\documentclass[reprint, amsmath,amssymb, aps, showkeys, onecolumn, notitlepage]{revtex4-2}

\usepackage{textcomp}
\usepackage{graphicx}
\usepackage{dcolumn}
\usepackage{bm}
\usepackage[mathlines]{lineno}
\def\thesection{\arabic{section}}
\def\thesubsection{\arabic{section}.\arabic{subsection}}
\def\thesubsubsection{\arabic{section}.\arabic{subsection}.\arabic{subsubsection}}
 
\setcitestyle{authoryear,open={},close={}} 
\begin{document}

\preprint{APS/123-QED}

\title{Road Network Evolution in the Urban and Rural United States Since 1900}
\author{Keith Burghardt}
 \email{keithab@isi.edu}
\affiliation{
 USC Information Sciences Institute, Marina del Rey, CA 90292, USA
}
\author{Johannes H. Uhl} 
\affiliation{Institute of Behavioral Science (IBS), University of Colorado Boulder, 483 UCB, Boulder, CO-80309, USA}
\affiliation{Cooperative Institute for Research in Environmental Sciences (CIRES), University of Colorado Boulder, 216 UCB, Boulder, CO-80309, USA}

\author{Kristina Lerman}
\affiliation{
 USC~Information Sciences Institute, Marina del Rey, CA 90292, USA 
}
\author{Stefan Leyk}
\affiliation{Institute of Behavioral Science (IBS), University of Colorado Boulder, 483 UCB, Boulder, CO-80309, USA}
\affiliation{Department of Geography, University of Colorado Boulder, 260 UCB, Boulder, CO 80309, USA}

\date{\today}

\begin{abstract}
Road networks represent a key component of human settlements, such as cities, towns, and villages, that mediate pollution and congestion, as well as economic development. However, little is known about the long-term development trajectories of road networks in rural and urban settings. We leverage novel spatial data sources to reconstruct and analyze road networks in more than 850 US cities and over 2,500 US counties since 1900. Our analysis reveals significant variations in the structure of roads both within cities and across the conterminous US. Despite differences in the evolution of these networks, there are commonalities and strong geographic patterns. These results persist across the rural-urban continuum and are therefore not just a product of accelerated urban growth. These findings refine and extend existing knowledge and illuminate the need for policies for urban and rural planning including the critical assessment of new development trends. 
\end{abstract}

                              
\maketitle
\section{Introduction}



Road networks are critical to local and (inter)national transportation and provide significant benefits to the economy, while also incurring significant construction and maintenance costs (in the hundreds of billions of US dollars) (\citet{Allen2019,Jaworski2019,Fraser2016}). The costs, however, can be mitigated by well-planned and maintained road networks (\citet{Fraser2016,Allen2019,Boeing2019efficient,Cervero1997}). The benefits of well-built road networks include greater walkability (\citet{Boeing2019efficient,Cervero1997,Gori2014}), which reduces environmental impacts, relieves public transport, increases transportation equity (\citet{Santos2008}), reduces travel time (\citet{Merchan2020}), and ultimately, improves city sustainability (\citet{Rao2018}). It is therefore crucial to both public health and the economy to understand how the road infrastructure has evolved, so as to learn about the effectiveness of past policies based on historical data.
However, historical road network data is scarce, impeding our quantitative knowledge about the past of road networks. Thus, researchers studying the longer-term evolution of road infrastructure typically rely on manually digitized road networks based on historical maps, which is labour-intensive and thus, often constrained to one or a few places (\cite{Masucci2013_london,Masucci2014_roadmodel,Kaim2020,Wang2019,Casali2019}). Only in recent years, larger-scale geospatial data integration efforts enabled the modeling and the analysis of historical road networks over larger spatial and temporal extents (\cite{boeing2020off,barrington2015century,barrington2020global}). Moreover, recent work leverages large amounts of (contemporary) road network data and applies advanced statistical and network-analytic methods to study the road network characteristics of cities and other spatial entities (\cite{Boeing2019efficient,Boeing2020multi,Barrington2019,Xue2021,Badhrudeen2022}). 
Specifically, researchers have studied how roads change over time using photography (\citet{Irwin2007}), integrating contemporary road network data with remote-sensing-derived data (\citet{barrington2020global}), with historical census tract data including information on residential structures (\citet{boeing2020off}), building construction year information (\cite{Fraser2016}), as well as cadastral parcel data containing building age information (\citet{barrington2015century}). Such efforts are also facilitated by the availability and accessibility of detailed, and highly complete contemporary road network data for many regions of the world, such as from OpenStreetMap (\cite{Boeing2917_OSMnx,Barrington-Leigh2017_complete}).
Alternatively, recent advances in computer vision and image processing enable the efficient automated extraction of historical road networks and other transportation infrastructure from historical maps, over large spatial extents (\cite{Jiao2021,saeedimoghaddam2020exploring,Uhl2022_maps,Hosseini2021}). These efforts contribute to an increasing availability of data on past road networks, enabling quantitative, multi-temporal analyses of road network change over large spatial extents.
In this vein, some recent work has explored the evolution of road networks in the United States (\citet{barrington2015century,boeing2020off}), and at the global scale (\cite{barrington2020global}), typically by analyzing topology-based network statistics constrained to portions of the contemporary road network attributed with a specific age estimate. These network statistics include measures of connectivity, geometric complexity, and network griddedness.
While these efforts provide unprecedented insight on the long-term trends of urban road networks during the 20th century in the context of urban sprawl (\cite{barrington2015century}) and highlight recent national-level trends (\cite{boeing2020off}), there are a few limitations, that this study aims to address. For example, the study of (\cite{barrington2015century}) focuses on connectivity-related aspects of road networks, and their analyses are limited to the urbanized parts of the U.S. (1990--2013), and to a subset of 10\% of U.S. counties (1920--2015), respectively. The work of (\cite{boeing2020off}) is spatially exhaustive, but has a focus on the griddedness of road networks, giving less attention to other aspects of road networks and their change over time.
Thus, existing approaches on road network evolution in the U.S. are either limited in their temporal range or geographic coverage, do not account for regional variation when performing longitudinal analysis, focus on specific aspects of road networks only, and do not address scale effects, manifested in the modifiable areal unit problem (MAUP) (\citet{Openshaw1979,Masucci2015}). 

This leaves a few important gaps in our knowledge: How have U.S. road networks evolved at fine spatial scales across an extensive time window? How do these evolution patterns vary regionally, between cities, and within cities? And how do different aspects of road networks change in relation to each other? Importantly, are the trends of road network evolution stationary across the rural-urban continuum, or are these trends dependent on the degree of urbanization? The latter point addresses the general tendency for urban planning literature to focus on urban areas at the expense of understanding rural and sub-urban settlements and development (\cite{Frank2014}). This lack of attention is unfortunate because accessibility is crucial to transportation policies, yet a focus on cities means limited knowledge of low-accessibility areas such as in periurban and rural settlements. Improving transportation equity and reducing financial hardships and pollution requires knowledge of all transportation regions including in rural settings.  

We aim to address these knowledge gaps with a nearly-exhaustive exploration of road network evolution across the conterminous U.S. (CONUS) since 1900. We explore the evolution of these networks from the fine-scale intra-city level up to the inter-metropolitan level through methods such as time series clustering and feature embedding. These methods provide insights into where changes in road network characteristics occurred and how factors such as topographic constraints or heavy population increase may impact these changes. Such associations remained hidden in previous, more aggregated analysis. We find that rural and urban areas experience similar patterns, which has not been fully appreciated in earlier work, including a reduction in the gridiron structure of newer road networks, a structure associated with more walkable neighborhoods (\citet{Boeing2019efficient,Cervero1997}). These results suggest common trends, such as the popularity of the automobile, contributed to this evolution. Moreover, significant differences in evolution also arise across the U.S., possibly due to differences in topography (e.g., mountainous and flat regions) or urban planning schools of thought.


Specifically, we analyze road network data from the 2018 National Transportation Dataset (\citet{NTD2020}), integrated with novel, spatial data layers containing historical built-up areas and building densities since approximately 1900 (\citet{uhl2021fine,leyk2018hisdac}). We reconstructed historical road networks under plausible assumptions that roads were built at roughly the same time as the oldest nearby houses (see Section 2.2) (\cite{boeing2020off,barrington2015century}). This integrated dataset enables us to study settlements and their changes through a road network lens at unprecedented temporal granularity and spatial resolution (\citet{Leyk2020,Uhl2021}). From these networks, containing in total over nine million nodes and over fifteen million road segments, we extract several road network statistics, such as the mean degree (the number of roads at each intersection), road density (the kilometers of road per unit area), a local griddedness metric, and the orientation entropy of road segments (\citet{Boeing2019efficient,Boeing2020world}), among others. These statistics can be used to quantify key characteristics of development within spatial units of different granularity (e.g., metropolitan areas, counties, grid cells). 


\section{Data \& Methods}

\subsection{Data}
\begin{figure*}[tb!]
    \centering
    \includegraphics[width=\linewidth]{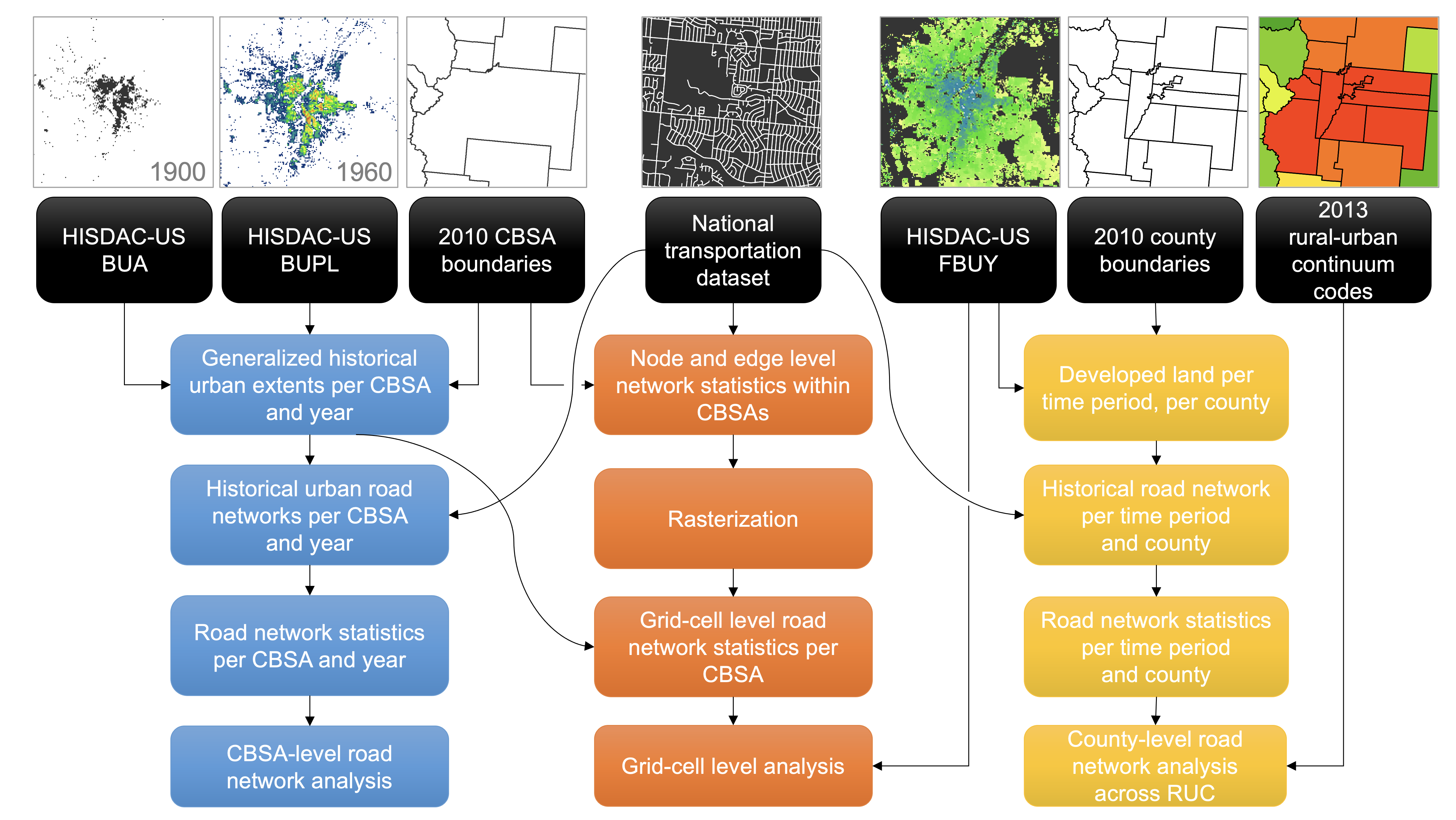}
    \caption{Input datasets and workflow of the three analytical components of this analysis at the CBSA, county, and grid-cell level.
    }
    \label{fig:Fig1}
\end{figure*}

\begin{figure*}[tb]
    \centering
    \includegraphics[width=\linewidth]{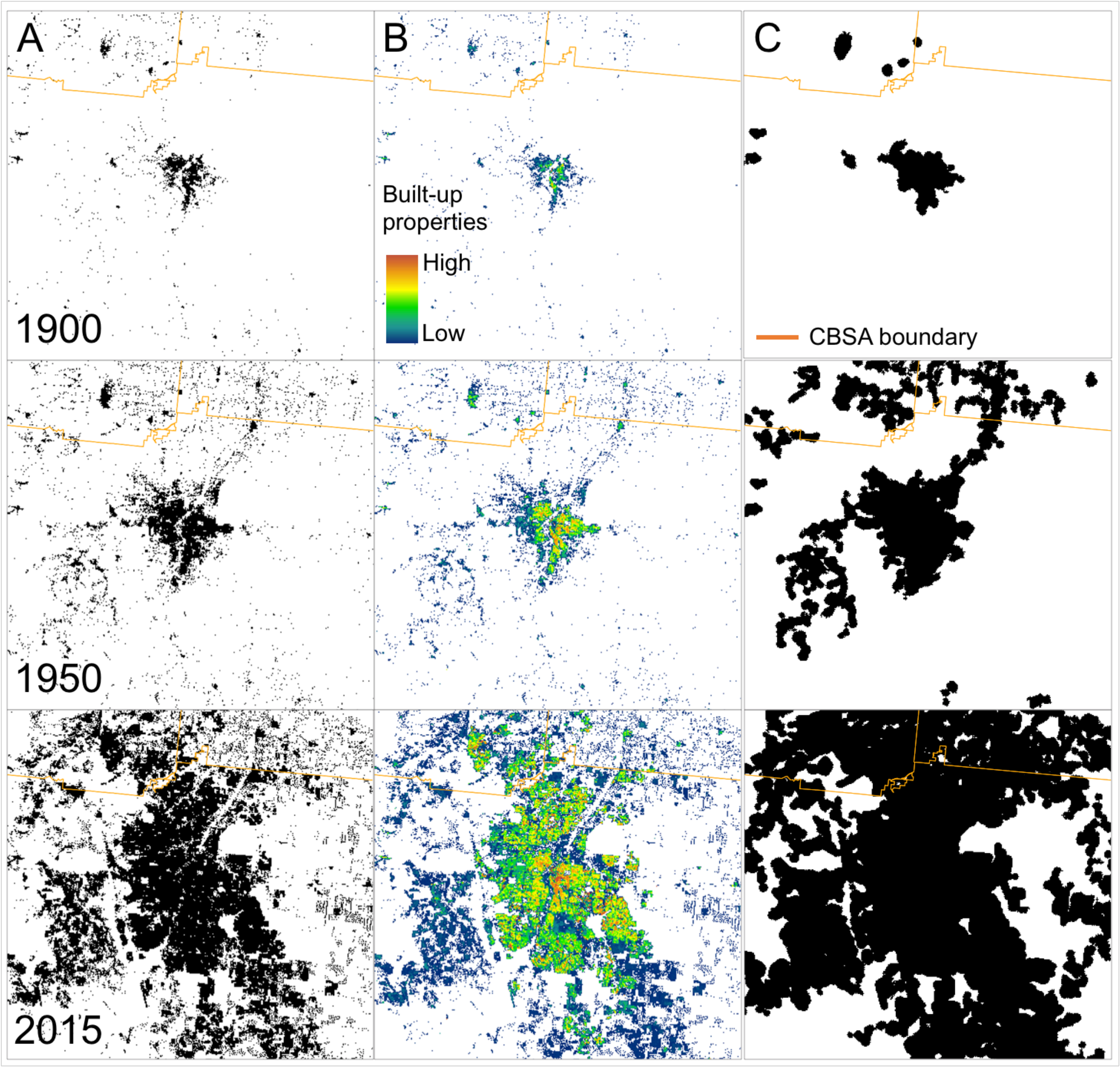}
    \caption{Creating spatially generalized, temporally consistent urban extents based on historical building density estimates. (a) Built-up areas from HISDAC-US, (b) Built-up property records (BUPL) as a proxy measure for built-up density, (c) density-based, spatially generalized urban extents, shown for Greater Denver (Colorado), all shown in 1900, 1950, and 2015.
    }
    \label{fig:Fig2}
\end{figure*}

Herein, we use geospatial vector data from the United States Geological Survey (USGS) National Transportation Dataset (\citet{NTD2020}), representing the US road network in approximately 2018. We model retrospective extents of built-up land with the Historical Settlement Data Compilation for the U.S. (HISDAC-US; (\citet{uhl2021fine,leyk2018hisdac})), which are derived from parcel-level built-year information contained in Zillow’s Transaction and Assessment Database (ZTRAX; (\citet{zillow2016})). More specifically, we use historical built-up areas (BUA) which are available in 5-year intervals from 1810 to 2016 as a series of binary, gridded surfaces at a resolution of 250m ((\citet{databua2020}), Figs.~\ref{fig:Fig1} \&~\ref{fig:Fig2}a). Likewise, we use historical estimates of the number of buildings per grid cell (built-up property locations; BUPL, Fig.~\ref{fig:Fig1} \&~\ref{fig:Fig2}b) (\citet{databupl2020}), as well as the First built-up year (FBUY), mapping the earliest year of development per grid cell (\citet{leyk2018hisdac,uhl2021fine}) (See Fig.~\ref{fig:Fig1}). While HISDAC-US data coverage is sparse in some rural areas of the US, geographic coverage and temporal information is largely complete in urban regions (\citet{uhl2021fine}), which we discuss in more detail in Section 3.4. Moreover, the accuracy of the built-up extents layer increases over time (\citet{leyk2018hisdac,uhl2021fine}), reaching acceptable levels after 1900 (\citet{uhl2021fine}). Thus, we constrain our analysis to the time period from 1900 to 2010. To measure road network characteristics for individual cities, we use the metropolitan statistical areas (MSAs) and micropolitan statistical areas ($\mu$SAs), defined by the US Office of Management and Budget, which allows results to be compared against previous work (\citet{barrington2015century,boeing2020off}). Collectively, these are known as Core-Based Statistical Areas (CBSAs) and roughly delineate cities based on the commuting patterns in their surroundings (\citet{CBSA}). MSA and $\mu$SA boundaries are shown in Supplementary Fig. S1. CBSAs nest within US county boundaries (\citet{CBSAinfo}). Counties are also employed in this study to model road network trends across the rural-urban continuum, using county-level rural-urban continuum codes (RUCC) provided by the US Department of Agriculture~(\citet{RUC}), classifying each county into one of nine levels of ``rurality'' (Supplementary Fig. S1).

\subsection{Methods}
Based on the various datasets, we develop a three-part analysis, aiming to assess long-term road network evolution at three different levels of spatial granularity including the analysis of (a) urban road networks at the city level, within CBSA boundaries, (b) road networks in urban, peri-urban and rural settings at the county-level, grouping them into strata of ``rurality'' by means of the county-level RUCC, and (c) intra-urban, local road network characteristics at the grid-cell level within CBSA boundaries. 

\subsubsection{Historical urban road network modeling}
We make a reasonable assumption that road networks remain unchanged in their geometry, once they are established (\cite{Scheer2001}), and that the evolution of road networks is largely characterized by expansion over time, and, to a lesser degree, by densification, which is in line with assumptions in previous work (\cite{boeing2020off,barrington2015century,Meijer2018}), due to the rarity of road network shrinkage. Changes in the geometric structure (e.g., layout, orientation) of road networks, or shrinkage are rare, and are assumed to be negligible in the case of the US during our study period. Thus, multi-temporal spatial data measuring the expansion of developed, or built-up land over time is commonly used to spatially constrain contemporary road networks to their assumed historical extents, under the assumption that the year of earliest settlement roughly corresponds to the year when nearby roads have been constructed (\citet{boeing2020off,barrington2015century,barrington2020global,Fraser2016}). 

Based on the gridded surface series BUA and BUPL from the HISDAC-US we develop an approach to generate spatially generalized urban extents, consistent across different cities and over time. In a first step, we generate a built-up density surface for each half-decade from 1900 -- 2010, within each 2010 CBSA boundary. To do so, we use circular focal windows of radius $r = 1$ kilometer, containing the proportion of built-up area within the focal neighborhood, derived from the BUA surfaces (Fig.~\ref{fig:Fig2}a). We then select all grid cells with a focal built-up density greater than 5\%. This method has previously been employed to discretize the rural-urban continuum into high density (urban) and lower density (peri-urban) strata (\citet{leyk2018assessing}) and shows high discriminative power between signals in remotely sensed spectral responses in urban settings. For each CBSA and year, we then segment the resulting contiguous groups (``patches'') of urban grid cells and compute the sums of built-up area, and number of buildings per patch (from the underlying BUPL surface, Fig.~\ref{fig:Fig2}b). We then compute the percentile ranks of the patches within a CBSA according to the number of buildings they contain. We discard small patches containing less than 10 buildings, likely representing scattered peri-urban settlements. To do so, we only retain patches that exceed the 90th percentile in the first year when the density filtering yields at least one patch of built-up land (which may be later than 1910 for late-developing cities). This way, we ensure that urban areas are modelled based on consistent criteria across space and time, and represented by smooth, contiguous, and largely gap-free areas (Fig.~\ref{fig:Fig2}c). We then clip the NTD road vector data to the urban delineations in each year, yielding sub-networks that can be uniquely identified by the combination of CBSA and year. We therefore model the intra-urban road networks for each CBSA and year, consisting of a total of 8 million nodes and over 10 million edges within CBSAs, as a basis to calculate a range of  road network metrics (Section 2.2.4).

\begin{figure*}[tb]
    \centering
    \includegraphics[width=\linewidth]{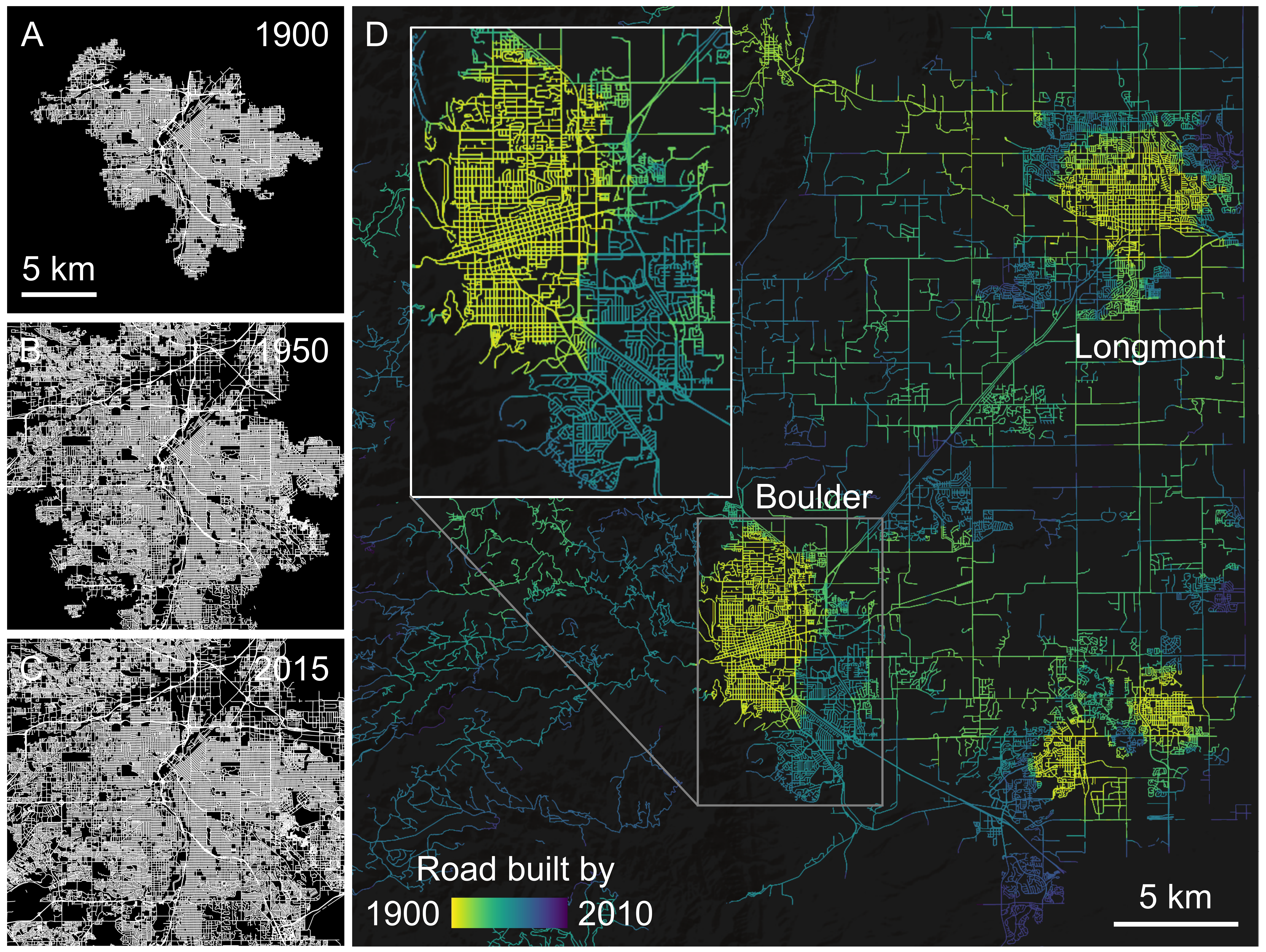}
    \caption{Growth of urban road networks over time. Denver metropolitan area in (a) 1900, (b) 1950, and (c) 2015; (d) Estimated road ages in a peri-urban setting near Boulder (Colorado) with street ages color-coded from light (1900) to dark (2015).
    }
    \label{fig:Fig3}
\end{figure*}

A detailed example of the reconstructed intra-urban road network is shown in Figs.~\ref{fig:Fig2} \&~\ref{fig:Fig3}a-c. Besides these binary data, we also attribute the road age estimate to each road network segment. The evolution of a subsection of the metropolitan area can be seen in Fig.~\ref{fig:Fig3}d, with lighter colors denoting older roads, and darker colors representing newer ones.

\begin{figure*}[tb]
    \centering
    \includegraphics[width=\linewidth]{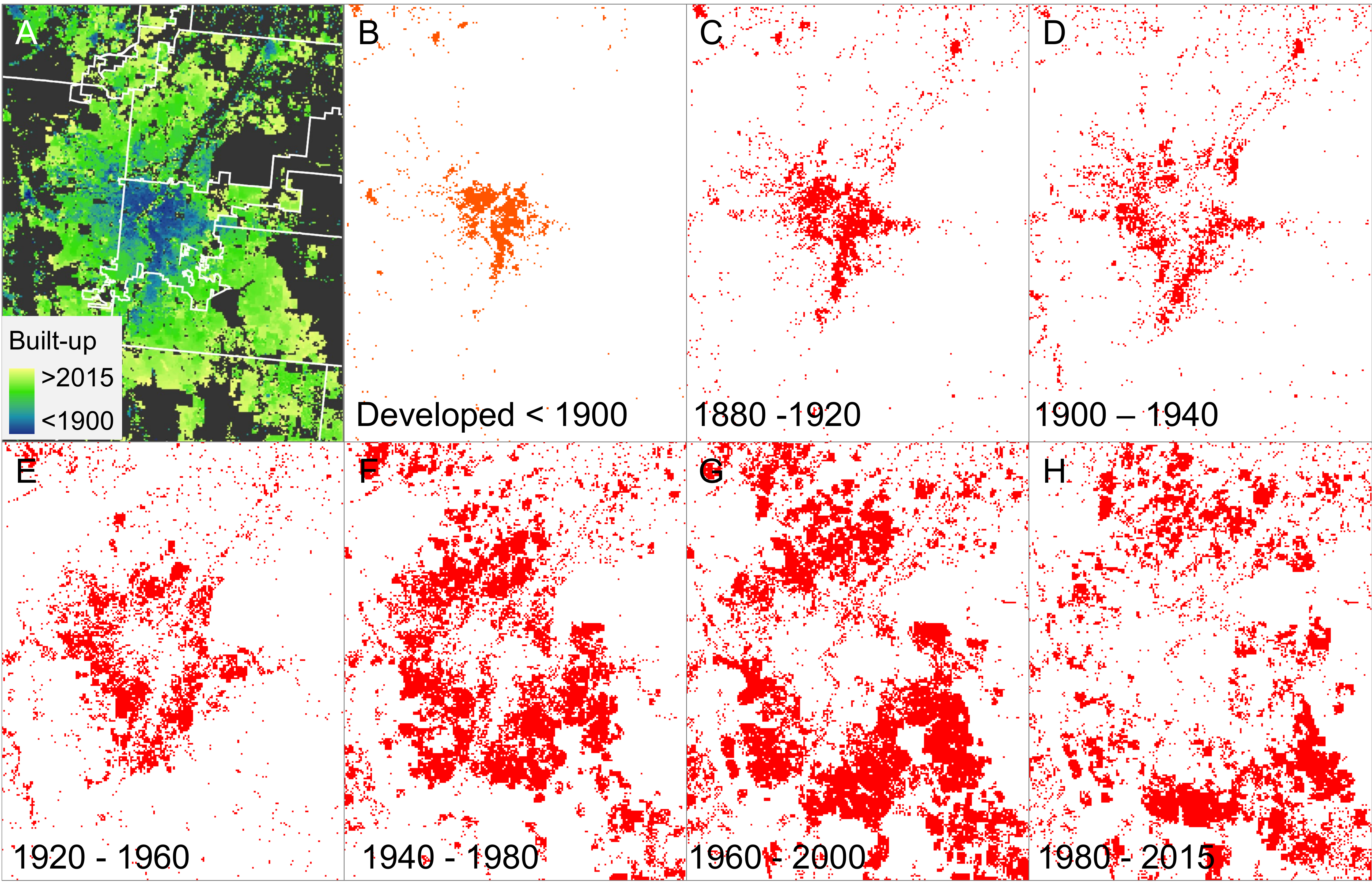}
    \caption{Spatial data layers generated to model historical road networks at the county level for different development periods. (a) the first built-up year (FBUY) surface from the HISDAC-US, indicating settlement age at the grid cell level, and (b) – (h) extracted areas developed within moving intervals of 40 years, with an overlap of 20 years, shown for Denver, Colorado. White lines in (a) represent county boundaries used as the spatial aggregation units. Historical road networks per county and development period are modeled by clipping the contemporary road network to the areas developed in each time period. 
    }
    \label{fig:Fig4}
\end{figure*}
\subsubsection{CONUS-wide historical road network modeling}
While the urban street networks we model in Section 2.2.1 allow us to characterize road network trends for large cities (metropolitan areas) and medium-size cities (micropolitan areas) over time, they do not cover the full urban-rural continuum, and are based on cumulative rather than incremental areal extents. To derive trends of road network characteristics over time and across the rural-urban continuum, we use the 2018 NTD road network vector data (\citet{NTD2020,RUC}) and the first built-up year dataset (FBUY), from the HISDAC-US data repository (\citet{leyk2018hisdac,uhl2021fine}). We first identify grid cells developed within moving temporal windows (i.e., time periods) of 40 years, shifted in steps of 20 years, e.g., developed prior to 1900, 1880-1920, 1900-1940, 1920-1960, etc., as shown in Fig.~\ref{fig:Fig4}. This is done to generate smooth trends avoiding abrupt changes in the extracted road network metric time series. For each county in the CONUS, we then extract the road network vector objects within the areas corresponding to each 40-year development period and assign an individual identifier to each contiguous group of developed grid cells (patches). We remove small, spatially isolated patches of under 0.31 square kilometers (corresponding to five 250 by 250 meter grid cells), as well as elongated patches of less than 500 meter width, likely representing settlements along highways and thus not relevant for characterizing road networks in cities, towns, or places (see Fig.~\ref{fig:Fig1} and Supplementary Fig. S2). For the remaining patches, containing over 27 million road segments, we calculate a range of road network metrics (see Section 2.2.4), aggregated per county and year. We analyze each of the network metrics in a bi-variate manner over time and across the rural-urban continuum, stratified by US census region (\citet{CensusRegions}), where the rural-urban continuum is based on the county-level rural-urban continuum codes (RUCC) provided by the US Department of Agriculture (\citet{RUC}), that classifies each county into one of nine levels of ``rurality.'' In total, we analyze 9.2 million nodes and 15.2 million edges across the rural-urban continuum. Both CONUS-wide and CBSA level historical road networks are extracted using ESRI ArcPy (\citet{ArcPy}) and Safe Software Feature Manipulation Engine Desktop (\citet{FME}).

\subsubsection{Gridded surfaces of road network metrics}
The road networks we extract in Sections 2.2.1 and 2.2.2 allow for the derivation of multi-temporal, road network statistics aggregated to the CBSA- and county-level. While these aggregation levels are expected to facilitate the quantification of trends across regional strata, or strata of different levels of rurality, they may ignore fine-grained spatial variations of road network characteristics within urban areas. Thus, we use grid cells of $1\times1$ kilometer as a third analytical unit for this study.

Using the FBUY gridded surface from HISDAC-US (\citet{uhl2021fine,leyk2018hisdac}), we calculate the average settlement age within $1\times1$ kilometer grid cells located within the 2015 urban delineations derived from the density-based delineation method described above using GeoPandas (\citet{Jordahl2020}) and SciPy (\citet{Pauli2020SciPy})) Python modules. The aggregation to $1\times1$ kilometer grid cells aims to avoid small sample sizes of road segments and intersections per grid cell, and thus, ensures the statistical support for the network statistics calculated per grid cell. We then identify all network nodes within a grid cell, as well as the centroids of all road segments (i.e., network edges) per grid cell. We calculate road network statistics for each grid cell based on the network statistics attributed to each node and to each edge (see Section 2.2.4). For consistency with the aggregated CBSA-level analysis, we calculate these grid-cell level statistics within the density-based generalized urban area from 2015 only (see Section 2.2.1).

\begin{table*}
    \centering
    \begin{tabular}{|p{3.5cm}|p{1.5cm}|p{4.5cm}|p{7.5cm}|}
 \hline
\textbf{Road network metric} & \textbf{Unit} & \textbf{Aggregated analytical unit} & \textbf{Description}\\\hline\hline
Degree & Node & grid cell, county, CBSA (average) & Number of roads touching a node\\\hline
Local griddedness & Node & grid cell, CBSA (average) & Number of quadrilaterals touching a node divided by its degree\\\hline
Road density & Edge	 &  Grid cell, county, CBSA & km road per km built-up area\\\hline
Orientation entropy & Edge & County, CBSA & Entropy of edge orientation angles, discretized into bins of $5^{\circ}$\\\hline
Azimuth variety & Edge & Grid cell & Number of unique edge orientation angles, discretized into bins of $5^{\circ}$\\\hline
Dead end rate & Node & Grid cell & Percentage of nodes of degree 1\\\hline
Nodes per km road & Node/edge & Grid cell & Number of nodes per km road within spatial unit\\\hline
Node density & Node & Grid cell & Number of nodes within spatial unit\\\hline
Percentage degree 4+ & Node & CBSA & Percentage of nodes with degree 4 or higher\\\hline
Road distance (total km road) & Edge & CBSA & Total km road within spatial unit\\\hline
Straight road rate & Street & County & Percentage of streets (i.e., edges with the same street identifier) with a edge orientation standard deviation $< 10^{\circ}$\\\hline
    \end{tabular}
    \caption{ Road network metrics used in this study.}
    \label{tab:Tab1}
\end{table*}

\subsubsection{Road network metrics}
For the historical road networks extracted per CBSA and year (Section 2.2.1) per county and development period (Section 2.2.2), and per $1\times1$km grid cell within CBSAs, we calculate a range of road network metrics (Table~\ref{tab:Tab1}). These metrics include the mean degree (the number of roads at each intersection), road density (the kilometers of road per unit area), measures of road network orientation (orientation entropy, azimuth variety), and several aggregated statistics, such as node density, nodes per km road, total road distance, the dead end rate, as well as a novel local griddedness metric. Many of these statistics are based on topology, but because the majority of junctions are straight there is less need to focus on non-trivial paths between intersections. This should not affect most metrics except entropy (which could be affected by non-trivial interactions with degree-two intersections) or griddedness metric, as we explain below. We use 1 km square grids to capture these statistics because less than 0.01\% of roads within CBSA boundaries are longer than 1 km. 

Griddedness has become critical to understanding walkability and related problems for cities in the US but measuring it has been difficult until recently. Namely, grid-like road networks appear to enhance walkability and lower relative vehicular travel in a city (\citet{boeing2020off,Cervero1997}). Griddedness (and related urban sprawl) metrics have been defined and implemented on several recent occasions (\citet{boeing2020off,Barrington2019,barrington2020global}). The methods to derive such metrics are sophisticated (\citet{boeing2020off}) and sometimes computationally expensive (\citet{Barrington2019,barrington2020global}). However, we aim for a simple, intersection-level measure to extend on previous work. 

We develop the local griddedness metric, a spatial complement to the clustering coefficient often used in network analysis (\citet{Watts1998,Boeing2020multi,Boeing2020world}). The local clustering coefficient of a node is defined as the proportion of triangles that exist whose vertex includes that node relative to the total number of possible triangles that could exist for a node of that degree. Unlike, e.g., social networks, road networks tend to be quadrilaterals, and more uniquely still, these cycles tend to be planar, meaning they all lie on a two-dimensional plane. 

These constraints offer guidance to a unique spatial clustering coefficient, local griddedness, which is the proportion of four-cycles containing that vertex relative to the total number of planar four-cycles for a node with that degree. Degree one, two, and three nodes are common and special cases for intersections, however. If a node is the end of a dead-end road, we define the local griddedness to be zero. We do not analyze nodes of degree two, because these nodes represent the continuation of a road, rather than an intersection. We show examples of this and explain our justification in more detail in the Supplementary Fig. S3. Finally, it is unlikely for a three-road intersection to have three city blocks meet there, but more likely is that it ends in a ``T.'' The maximum number of city blocks is therefore defined as two and is otherwise the degree. While in most cases, local griddedness is between 0 and 1, we allow for rare instances in which, for example, degree-three nodes have a value up to 3/2 (a ``super gridded'' node), for the T intersection to have a natural griddedness value of 1.0. This also means that roads that violate this planar assumption (e.g., those with bridges) may be greater than 1, but such instances are rare. Using this measure, any node can have its griddedness value rapidly calculated, allowing for extremely fine-grained analysis of road network spatial statistics. 

Because this metric is based on the road network topology, a limitation of this technique is that some nodes can be misidentified as having a high griddedness value, although this is rare from visual inspection. Topology is useful both because it appears reasonable and is very fast, which partly motivates its use in, e.g., (\cite{Figueiredo2007, Caldarelli2004}). Our method differs from previous work because it is node-centered, instead of edge-centered (\cite{Figueiredo2007}), and is normalized such that values near 1.0 are irongrid-like, while some similar metrics are not as easy to interpret (\cite{Caldarelli2004}).

In addition to this metric, we separately calculate azimuth variety and orientation entropy by binning the angles between road intersections into six-degree wide bins. Changing the width of the bins does not qualitatively change our findings but can change the absolute value of entropy (which can be as high as the log of the number of bins) as well as the azimuth variety.

Finally, we quantify the proportion of cul-de-sacs (i.e., the dead end rate). This metric requires that end points of the road vector lines that are introduced by the clipping are recorded and excluded from subsequent node analysis because these nodes represent artificial cul-de-sacs introduced by the data processing that would yield inflated dead end rate values.
The road network metrics at CBSA and grid cell level are computed using NetworkX, and the statistics per development period and county are obtained using Safe Software Feature Manipulation Engine.

\begin{figure*}[tb]
    \centering
    \includegraphics[width=\linewidth]{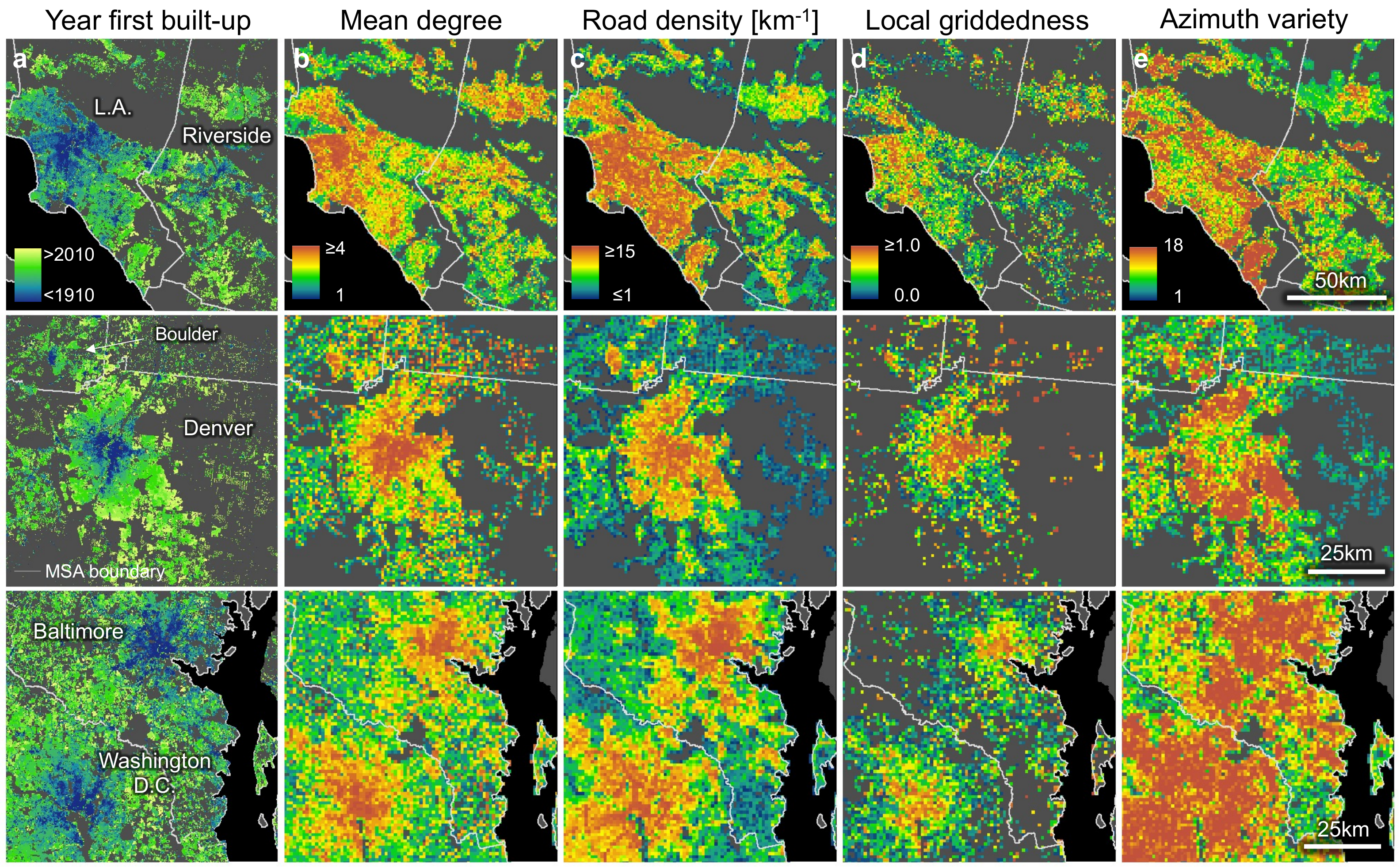}
    \caption{Gridded ($1\times1$ kilometer) road network statistics within cities. Columns, left to right: inferred date of road construction, mean degree (number of roads at each intersection), road density (kilometers of road per square kilometer), local griddedness (degree to which an intersection is part of a grid topology), and azimuth variety (variety of road orientations, discretized to 10$^{\circ}$ bins). Rows represent from top to bottom: Los Angeles, Denver, and Baltimore-Washington metropolitan areas.
    }
    \label{fig:Fig5}
\end{figure*}

Figure~\ref{fig:Fig5} demonstrates how these statistics reveal interesting differences in road network structure between, as well as within, metropolitan areas. For example, azimuth variety (i.e., the variety of unique road orientation angles) is high (representing very irregularly oriented roads) in the Baltimore Washington area and low (regular roads) in the periphery of the Denver and Los Angeles metropolitan areas. Similarly, strong variations can be seen for the local griddedness metric. We observe in Fig.~\ref{fig:Fig5} low local griddedness (fewer square blocks) and low edge density (spaced out roads)in some areas of the metropolitan areas of Los Angeles, Washington DC, and Denver where newer roads were built.

\subsubsection{Metropolitan-level historical road network statistical analysis}
To construct CBSA-level analyses of the road networks, we group the patches constructed in Section 2.2.1 into CBSA regions. If a patch is on the border of a CBSA region, we cut it off at the boundary, and edges that reach the boundary are removed; this is not common but is a reasonable way of defining patches associated with only one metropolitan or micropolitan area. At ten-year intervals between 1900 and 2010, as well as for 2015, we construct the road network topology using the Python library NetworkX, and remove nodes with degree two from the analysis, which we explain in more detail in Section 2.2.3 and Supplementary Fig. S3.  For each CBSA, we record the total road length, area, degree, proportion dead ends and degree greater than or equal to four, as well as orientation entropy and local griddedness. These raw statistics are used to construct combined measures, e.g., the distance per unit area to be able to quantify the road length distance within all patch areas (which are a small proportion of the total CBSA area). 

\subsubsection{Grid-cell-level correlation analysis and time series clustering}

In total, we calculate seven grid cell-level network statistics (cf. Table~\ref{tab:Tab1}). Due to potentially small sample sizes within grid cells, we replace the orientation entropy by the variety of unique azimuth values per grid cell, calculated after discretizing the road segment azimuth into bins of ten angular degrees. Based on the gridded surface indicating the average age per grid cell, and corresponding cell-level network statistics within each CBSA, we extract cell-by-cell pairs of settlement age and road network statistics for each city. These vectors enable us to calculate correlations between age and network characteristics, for each city, considering the local, fine-grained variability of settlement age as it is associated with the characteristics of the road network. Moreover, we generate time series of each network characteristic for each city. To characterize the relative relationship between age and network characteristics, we discretize the age surface per CBSA into deciles. Thus, the resulting time series consists of the same number of observations (i.e., ten) and are independent from the absolute age of the cities. For each of the network characteristics, we conduct time series-based cluster analysis separately for MSAs and $\mu$SA. We use the time-series $k$-means algorithm, TSK-means (\citet{Huang2016}),  in conjunction with the Dynamic Time Warping (\citet{DTW}) similarity metric to characterize the dissimilarity between time series, implemented in the tslearn (\citet{Tavenard2020}) Python module. In order to identify the optimum number of clusters $k$, we calculate the cluster inertia based on DTW similarity as a measure of separation between time series clusters for a range of $k$ from two to twenty and use the popular elbow method (\citet{Syakur2018}) to identify the approximate number of clusters for each scenario. We normalize the cluster inertia of each clustering scenario into the range (0, 1) to compare the cluster quality across the different network statistics. Moreover, we assess the agreement of the CBSA clusters identified for different road network metrics using Normalized Mutual Information (NMI) (\cite{Forbes1995}).

\section{Results}
We carry out longitudinal and cross-sectional studies of the evolution of the US road networks since 1900, at spatial scales ranging from the grid cell level to the CONUS. Firstly, we present regional trends of urban road network evolution (Section 3.1), and identify types of city-level road network evolution (Section 3.2). Finally, we present the county-level trends of road network evolution across different development periods and across the rural-urban continuum (Section 3.3).

 We begin our analysis on CBSA-level trends.

 \begin{figure*}[tb]
    \centering
    \includegraphics[width=\linewidth]{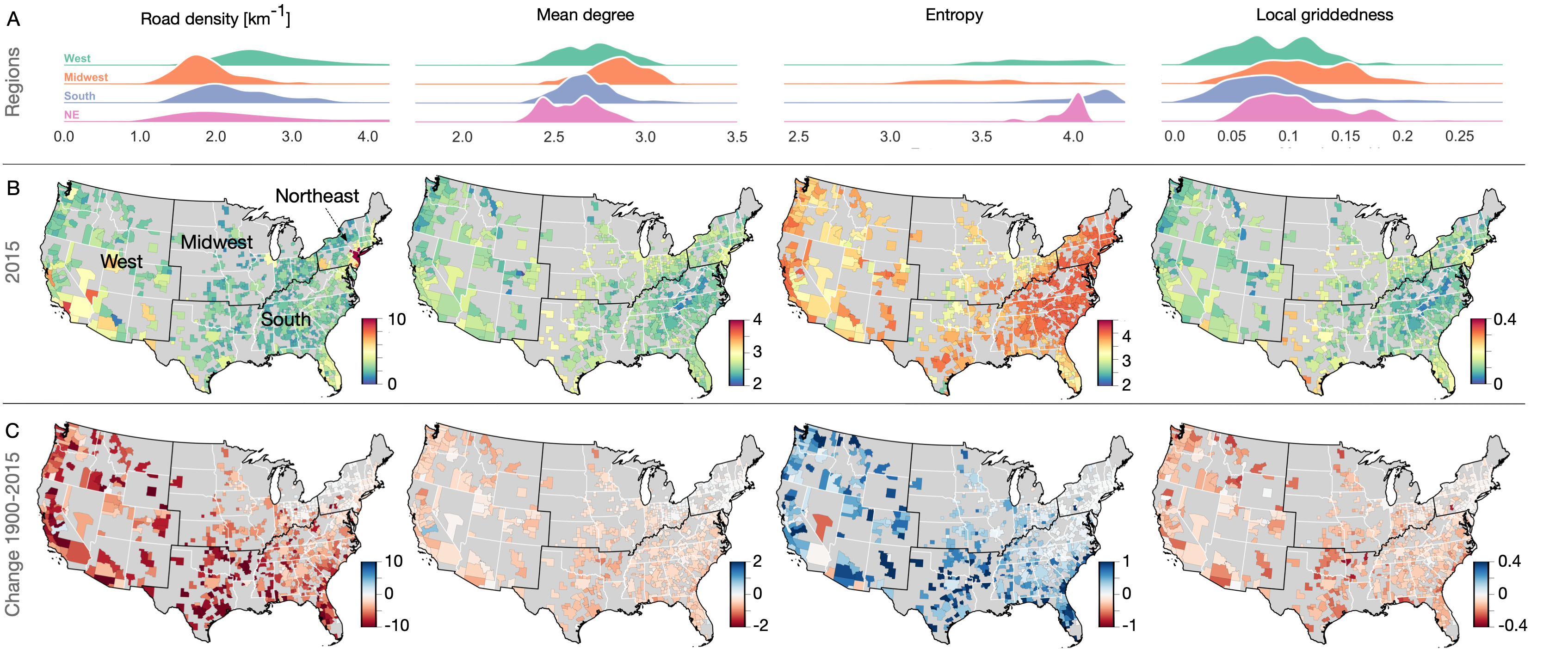}
    \caption{Variations in road network statistics across the CONUS. (a) statistical distributions for road density, mean degree, orientation entropy (\cite{Boeing2019efficient}), and local griddedness, split by region. (b) Road statistics for each CBSA. (c) Changes in road network statistics between 1900 and 2015.
    }
    \label{fig:Fig6}
\end{figure*}

\subsection{Regional trends of road network evolution}
At the national level, Fig.~\ref{fig:Fig6} reveals broad trends in how road density varies across U.S. census regions ((\citet{CensusRegions}); henceforth referred to as regions). We find relatively low road density in the Midwest as well as the Northeast although the distribution in the Northeast shows a pronounced broad tail due in large part to the New York City MSA (Fig.~\ref{fig:Fig6}a). The mean degree (i.e., mean number of roads per intersection) of networks, in contrast, is typically higher in the Midwest and lowest in the more mountainous regions (e.g., the Appalachian Mountains) near the East coast (Fig.~\ref{fig:Fig3}b). In agreement with these statistics, we find that orientation entropy, a proxy of the road network’s regularity, is lowest (most regular) in the Midwest and highest (least regular) in the South and Appalachia. Complementing these observations, local griddedness and mean degree are highest in the Midwest and lowest in the South and mountainous regions in the West. Our temporal analysis, meanwhile, reveals that across 115 years some regions, such as the South and West, have seen great changes in their road networks, while the Northeast has been relatively stable due to limited additions of new roads in an already developed region (see Fig.~\ref{fig:Fig6}c and Supplementary Fig. S4). In general, however, newer road networks tend to be less grid-like and less densely packed as they expanded into suburban areas (Fig.~\ref{fig:Fig3}).

\begin{figure*}[tb]
    \centering
    \includegraphics[width=\linewidth]{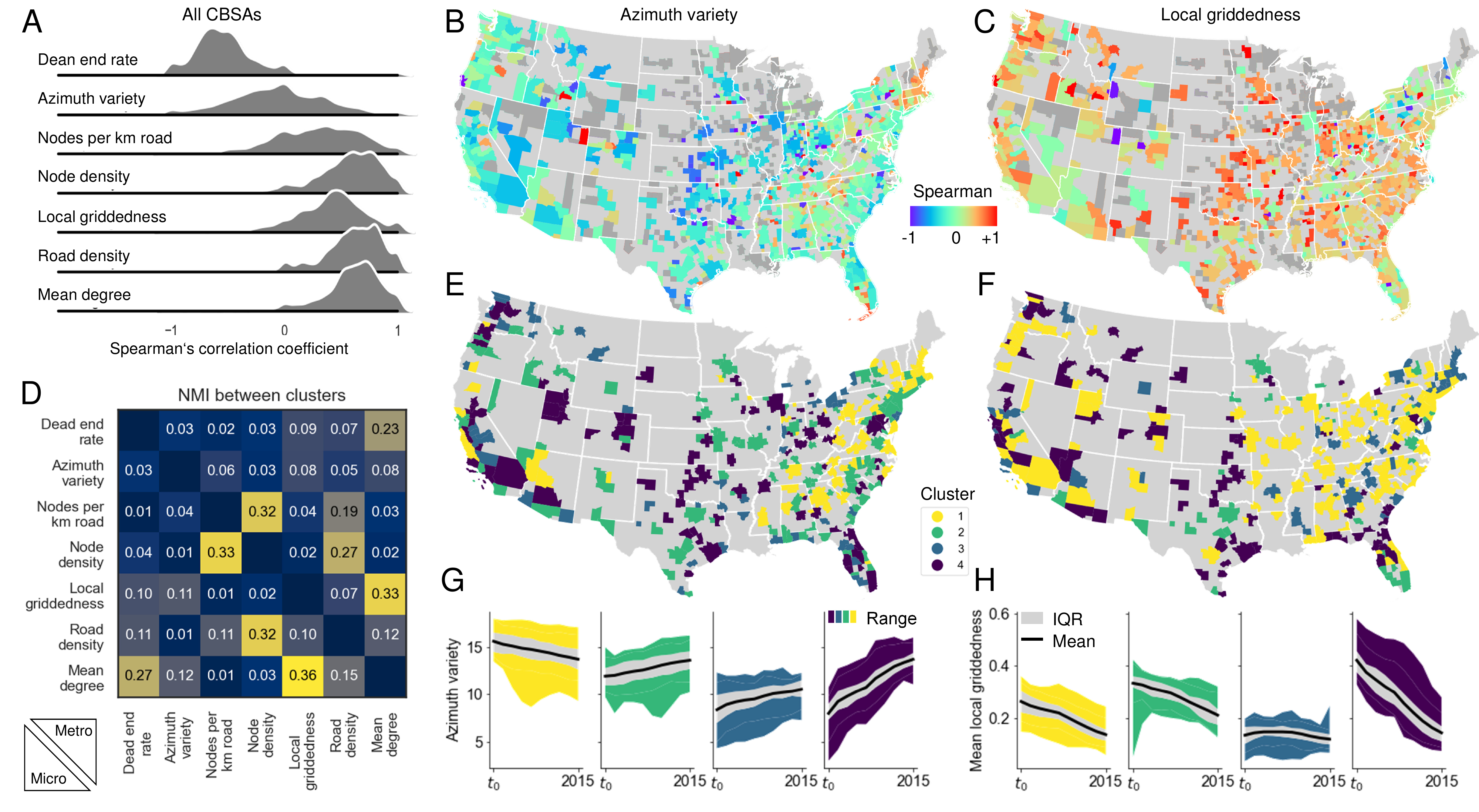}
    \caption{Grid-cell level correlation and time series clustering. (a) Distributions of CBSA-level Spearman correlation between network statistics and settlement age. (b–c) maps of Spearman correlation between age and azimuth variety and local griddedness, respectively, for CBSAs. (d) Normalized mutual information (NMI) between time series k-means-based clusters for each network statistic. (e–f) Spatial distributions of k-means-based time series clusters with $k=4$ for azimuth variety and local griddedness in MSAs. (g–h) The time series of the clusters found in (e–f).
    }
    \label{fig:Fig7}
\end{figure*}

\subsection{The evolution of road networks at the city-level}
Fig.~\ref{fig:Fig7} illustrates the evolution of road networks within metropolitan areas based on network statistics extracted for each $1\times1$ kilometer grid cell, since the first recorded building was constructed. For this analysis, we use several metrics including the ratio of dead end roads, the number of intersections per kilometer of road, the number of nodes, and length of road per area unit, and mean degree. We assess correlation of these metrics with the age of each grid cell, as shown in Fig.~\ref{fig:Fig7}a (temporal correlations for all metrics are shown in Supplementary Figs. S5 \& S6). Broadly speaking, density and griddedness-related metrics decrease over time, while azimuth variety shows mixed trends during the study period, and the dead end rate increases with road network age. These trends are similar for large cities (MSAs) and smaller cities ($\mu$SAs; see Supplementary Fig. S6). Figure~\ref{fig:Fig7}b–c shows the spatial distributions of correlation coefficients between age and the two metrics azimuth variety and local griddedness, respectively, revealing strong spatial patterns. We try capturing this spatial variation by computing k-means clusters of temporal patterns in these statistics (\cite{Huang2016}). We measure similarity between the time series of the CBSAs using the Dynamic Time Warping (DTW) distance metric (\cite{DTW}). This metric yields large distances for time series that differ considerably in their trend, shape, and/or timing. We find the data series can be grouped well into just 3 clusters (for MSAs), and 4 clusters (for $\mu$SAs), as indicated by the ``elbow'' in the DTW-based cluster inertia in Supplementary Fig. S8. We separately analyze MSAs and $\mu$SAs to understand how their evolution is affected by the size of the urban area, as CBSAs of similar age can evolve very differently. 
The agreement of these CBSA clusters based on individual road network metrics (measured by the Normalized Mutual Information (NMI) (\cite{Forbes1995}) varies between metrics, but is consistent across MSAs and $\mu$SAs (Fig.~\ref{fig:Fig7}d). High NMI values as observed for pairs of metrics such as mean degree and local griddedness indicate that CBSAs are separated into clusters in a similar manner, indicating that the temporal trajectories of the road network griddedness and mean degree follow similar evolution types. When mapping the computed clusters as shown in Fig.~\ref{fig:Fig7}e–f, we find that the identified ``types'' of road network evolution at the city level follow strong spatial patterns. For example, we find that nearby cities in the Appalachian region or in the Northeast have similar trends in their azimuth variety. These results reveal a Simpson’s paradox (\citet{Simpson1951}) in that the trends in disaggregated data differ from the overall trends shown in Supplementary Fig. S7 and in previous work (\citet{boeing2020off,barrington2015century}). 
Griddedness trajectories for MSAs, however, are distinct between the East and the West. The temporal patterns for these clusters are shown in Fig.~\ref{fig:Fig7}g--h, where azimuth variety grows fast in MSAs of clusters 1,2, and 3, and decreases for cluster 4 (roughly covering Appalachia and the Northeast) where starting values were highest. In contrast, griddedness in MSAs decreases most in cluster 4 (West, Midwest) where values are highest and slower elsewhere (including the coastal regions). The clustering results for all seven grid-cell level statistics are shown in Supplementary Fig. S8.

\begin{figure*}[tb]
    \centering
    \includegraphics[width=\linewidth]{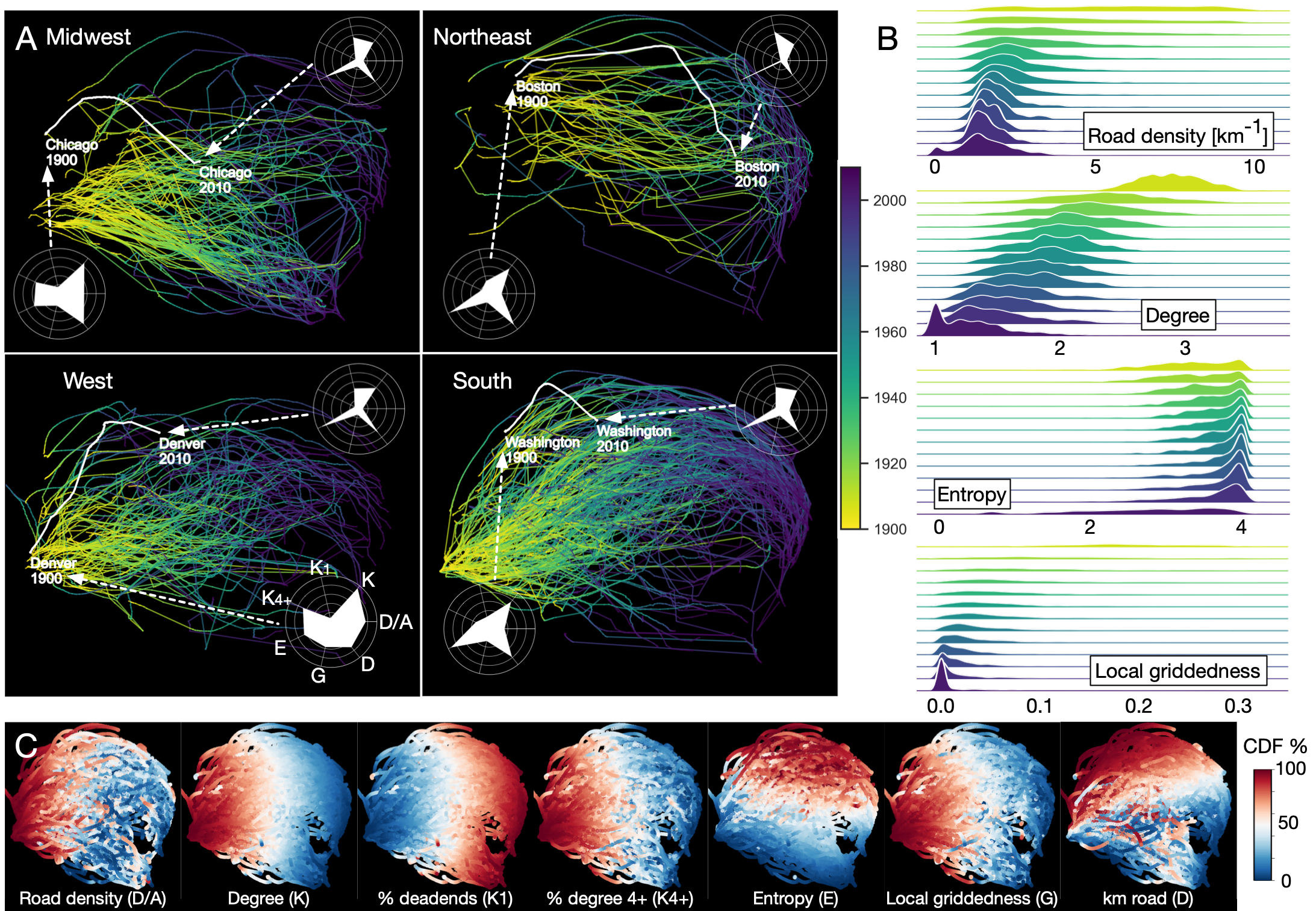}
    \caption{Road network statistics at the CBSA-level over time. (a) UMAP plot of cities embedded into seven road network statistics for each US region. Radar charts are road network statistics for example MSAs whose outer ring corresponds to the largest value of each network statistic. (b) Distributions of road density, mean degree, entropy, and mean griddedness per CBSA over time. (c) A closer inspection of the UMAP plots in (a), color-coded by cumulative distribution of mean city statistics from 0\% (low values compared to other cities) to 100\% (higher than 100\% of other cities). These figures reveal network statistics vary strongly, such that cities in the upper left-hand corner of each plot have many roads built per year and have high entropy. Networks in the lower left-hand corner have higher local griddedness, road density, \% degree 4+, and average degree, as well as lower \% dead ends. 
    }
    \label{fig:Fig8}
\end{figure*}

While these univariate trends provide interesting insight, how do metrics for each city vary over time? The multivariate trajectories of CBSA-level network statistics over time are visualized in Fig.~\ref{fig:Fig8}a, where we embed statistics for each city into two dimensions using UMAP (\citet{McInnes2018}). UMAP is a more nuanced version of PCA embedding, where in this lower-dimensional space the relative position of datapoints are approximately preserved. The embeddings are based on seven statistics computed for each CBSA and smoothed over time: the proportion of dead ends, mean degree, road distance per area, log of road distance, local griddedness and entropy (\cite{Boeing2019efficient}), and proportion of intersections with four or more roads (details on data smoothing are seen in Supplementary Figs. S9 \& S10). Changes in statistics are highlighted by radar charts computed for the Chicago, Washington, DC, Boston, and Denver MSAs. In Fig.~\ref{fig:Fig8}c, we also show how these statistics vary across the UMAP projection, thus providing insight into the trends of individual cities, in a way similar to Badhrudeen et al., (\cite{Badhrudeen2022}). Our results demonstrate broad similarities but also considerable variation in city-level trends over time. Cities across but also within regions differ in their routes to their final statistics, yet again pointing to Simpson’s paradox in our data: trajectories, disaggregated to the level of a metropolitan region can differ, sometimes substantially, from any assumed overall trend, regionally or nationally. Nonetheless, we see some trends are consistent across cities, as shown in Fig.~\ref{fig:Fig8}b, such as lower road density, fewer roads per intersection, and statistics consistent with less gridiron-like roads (although orientation entropy has recently started to decrease, possibly implying more regular angles between intersections). Results are robust to data cleaning (Supplementary Fig. S7). These plots illustrate the heterogeneity in the evolution of cities that resulted in today’s urban areas of the US. 

Figure~\ref{fig:Fig8}b, meanwhile, reveals changes in road network statistics over time. There are general trends of increasing entropy and decreasing road density and griddedness across the CONUS. We observe significant variance in the computed statistics across cities in early times, especially with road density and local griddedness. However, there are also notable commonalities, such as a tendency for newer regions to have lower griddedness and higher entropy (although entropy’s trend is non-linear). The trends for mean degree and mean griddedness reveal decreasingly grid-like networks over time (statistical significance of results are shown in Supplementary Fig. S11). While scholars have argued that grid structures enable efficient traffic flows and thus, may contribute to reduce emissions, congestion, and to increase the use of alternative, environmentally friendly transportation methods (\citet{Boeing2019efficient,boeing2020off,Cervero1997,Sharifi2019,Gao2013}), more recently developed road networks appear to be less effective in that regard. But why? Some trends may be due to urbanization expanding into hilly topographies, such as in the mountains north of downtown Los Angeles or the Piedmont region of western Maryland (see Fig.~\ref{fig:Fig2}), where grid-like road networks and high road densities are not feasible. Residential development in these topographically more complex areas but also altered development patterns in periurban areas may help explain the decline in urban densities (\cite{Gao2013,Angel2017}), but future work needs to analyze these hypotheses in greater detail.

\begin{figure*}[tb]
    \centering
    \includegraphics[width=\linewidth]{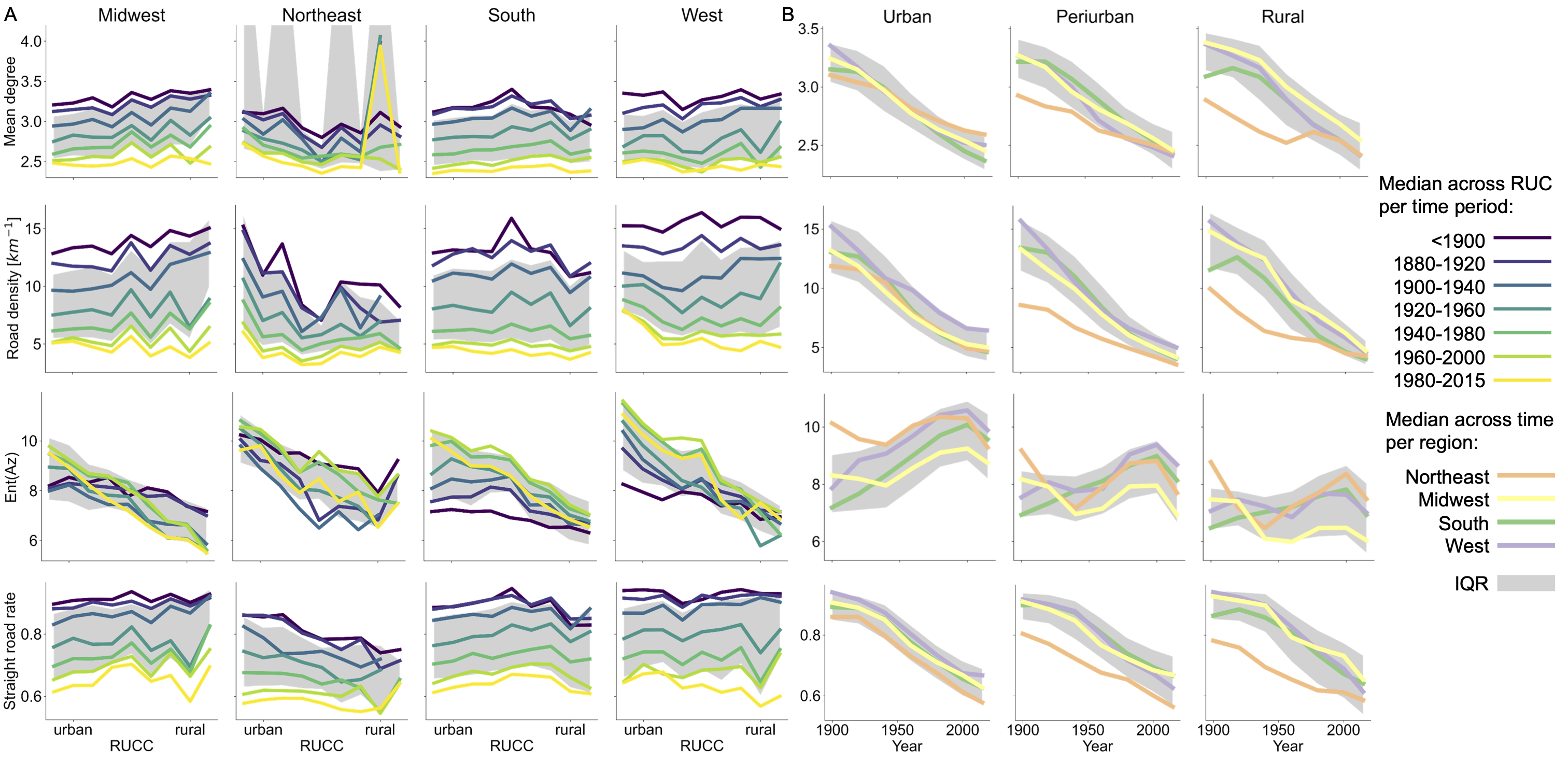}
    \caption{County-level network statistic trends across time, regions, and the rural-urban continuum. (a) Road network statistics for different time periods, as a function of the urban-rural gradient (defined by county-level RUCCs) for each CONUS region. (b) Corresponding visualization of these decomposed trends over time, aggregated by region. Solid lines are the median values of county-level statistics within the strata defined by RUCC, time period, and region.
    }
    \label{fig:Fig9}
\end{figure*}

\subsection{Road networks across the rural-urban continuum}
Finally, we analyzed and compared trends along the full rural-urban continuum (including counties outside of the CBSAs), using county-level rural-urban classes (\cite{RUC}), stratified by regions (rural-urban continuum values for each county are shown in Supplementary Fig. S1). While significant insight has been gleaned from analysis of urban road network growth (\cite{barrington2015century,boeing2020off}), Fig.~\ref{fig:Fig9} reveals its complement, road network growth in periurban and rural settlements. These results demonstrate that our findings generalize to rural areas. Namely, the mean degree, road density, and straight road rate in developed areas within rural and urban settings are all decreasing from 1900 to 2015 across all regions within the CONUS. Differences in statistics across the rural-urban continuum are often, however, statistically significant (Supplementary Fig. S11 \& S12). Trends vary most for orientation entropy, which tends to increase in urban settlements, but is stagnant and low in rural settlements. Comparing the road networks established in a fixed time across the rural-urban continuum, we observe high levels of persistence, indicating similar road construction trends in urban and rural places in a given time period. One notable exception are rural settlements in the Midwest, which initially have a higher road density than urban settlements, but this reverses for roads built in more recent time periods, and therefore looks more like the rest of the US.

While we focus on rural and urban trends in this section, the observed trends are also consistent when split by MSA or $\mu$SA, region, or year of city’s maximum development (see relation between maximum development year and regions in Supplementary Fig. S13), with some minor differences, as shown in Supplementary Fig. S14.

\subsection{Uncertainty, validation and sensitivity analysis}
While the HISDAC-US data provides accurate building ages across the US, these data are incomplete. More specifically, many buildings are missing their build year (temporal incompleteness) and some are missing from the dataset entirely (incomplete geographic coverage). We quantified the geographic coverage by comparing the buildings found against the Microsoft Building Footprint dataset (\citet{MBF2020}), and the temporal completeness by quantifying the availability of YearBuilt in ZTRAX (Supplementary Figs. S15 \& S16). We also assessed the sensitivity of our main results to different levels of completeness, by systematically excluding CBSAs of lower completeness levels. Our findings are robust to such variations in data quality (see Supplementary Figs. S17, \& S18). 


 %

To further verify our results, we compare some of our findings to previous work (\citet{boeing2020off,Boeing2020multi,barrington2015century}). We compare differences in statistics between small and large cities (Supplementary Fig. S19) and regional network statistics (Supplementary Fig. S20), as well as trends based on most statistics, such as mean degree or dead-end rate (Supplementary Figs. S21), and all statistics in the present study broadly agree with previous research. However, in contrast with recent research (\citet{boeing2020off,barrington2015century}), we did not find a significant increase in mean degree or proportion of four-way intersections in the 21st century. Some of these results are likely data-dependent (\citet{barrington2015century}) but some observed differences may also be due to the MAUP (\citet{Openshaw1979,Masucci2015}). Our large dataset allows us to analyze historical road networks, which are reconstructed based on the age information of nearby settlements, at finer scales that do not depend on pre-defined boundaries at coarser resolution, such as a census tract. 

\section{Conclusions}
We demonstrate how integration of large spatio-temporal datasets enables new detailed insights into long-term evolution of human settlements through the lens of road networks across the rural-urban continuum in the US. We measure road network characteristics over time within varying units of analysis and differentiate resulting trajectories across geographical regions. This data-driven approach reveals regional patterns that fill important knowledge gaps in our understanding of how road networks have evolved, possible drivers of these changes, and what kind of differences we find in these networks across cities and regions. The continuous reduction in the proportion of gridiron roads is of particular importance as this reduction is associated with reduced walkability of neighborhoods (\citet{barrington2015century,boeing2020off}), which contrasts with the popular New Urbanist school of thought (\citet{barrington2015century}) that promotes walkability of cities. Our findings notably reveal similar trends in rural regions which have been neglected in previous research. This is somewhat unexpected due to the persistently low population density in rural settings and suggests a reflection of existing policies and concepts is needed to promote greater neighborhood walkability. 

The presented findings can offer a new understanding of the importance of various network characteristics over space and time and thus shed light on the various forms of development during different time periods and across regions. We specifically address under-explored differences in growth patterns across and within urban areas, which were hidden in aggregated data, and analyze the growth across the rural-urban continuum. These results are analyzed across a larger time span, from 1900 when automobiles were rare to 2015, when they were ubiquitous and changed the urban infrastructure landscape. The patterns, such as in Fig.~\ref{fig:Fig7} are distinct from those expected in aggregated data, a property known as Simpson's paradox. Finally, our insights could help policymakers better understand the (un)intended impacts of infrastructure development, both now and in the past, to inform future planning efforts. For example, past work has discussed how many highways were built at the expense of minority neighborhoods (\citet{Fitzpatrick2000,Karas2015,Mohl2004}), but there is relatively limited work quantifying the effect this, and other infrastructure projects, had on minority communities (\cite{Houston2004}). While some previous research discussed pollution in rural neighborhoods (\cite{Houston2004}), understanding the potential pollution impact long into the past has been lacking. Furthermore, the relative impact of infrastructure on communities in different cities is still under-explored. The development of road infrastructure can also offer economic benefits that have yet to be fully quantified within urban areas (\citet{Jaworski2019}) as well as rural settings. Given the high priority of infrastructure investment planning in the US, these insights are of particular importance and need to be considered an integral part of urban and rural planning. 

Our work can also help researchers uncover the mechanisms that drive road evolution (\citet{Zhao2016,Barthelemy2008,Masucci2014_roadmodel,Strano2012,Bettencourt2007,Barthelemy2013}) by comparing the spatial statistics predicted in models to those seen in the data. These models predict, for example, how the mean road network degree and degree distribution varies as a function of city size, which can be tested with the cities we analyze. Our results already show how these models can be further improved, such as by accounting for the regular grid pattern of older cities, which is distinct from that expected by Barthelemy and Flamini (\citet{Barthelemy2008}). These differences can help researchers understand where their mechanistic assumptions differ from data and improve our understanding of what drives urbanization, locally and regionally. 

While our results point to significant variation as well as commonalities in road network evolution, further analysis is needed to understand the global evolution of these networks (\citet{barrington2020global}), and their trends over extended periods of time. Moreover, the presented study focuses on local roads within developed areas. Thus, future work will also include highways that connect the local road networks. Moreover, these results only approximate the network existing at a given time, as our models focus on network growth, and ignore network shrinkage (i.e., roads disappearing over time). Future work needs to therefore explore the historical network through, for example, automated analysis of historical maps (\citet{saeedimoghaddam2020exploring,Uhl2022_maps,Jiao2021}) or other records (\citet{Erath2009}).

\clearpage
\section*{Supplementary Materials}

\setcounter{figure}{0}
\renewcommand\thefigure{S\arabic{figure}}  
\vspace{100pt}
   \begin{figure*}[tbh!]
    \centering
    \includegraphics[width=\linewidth]{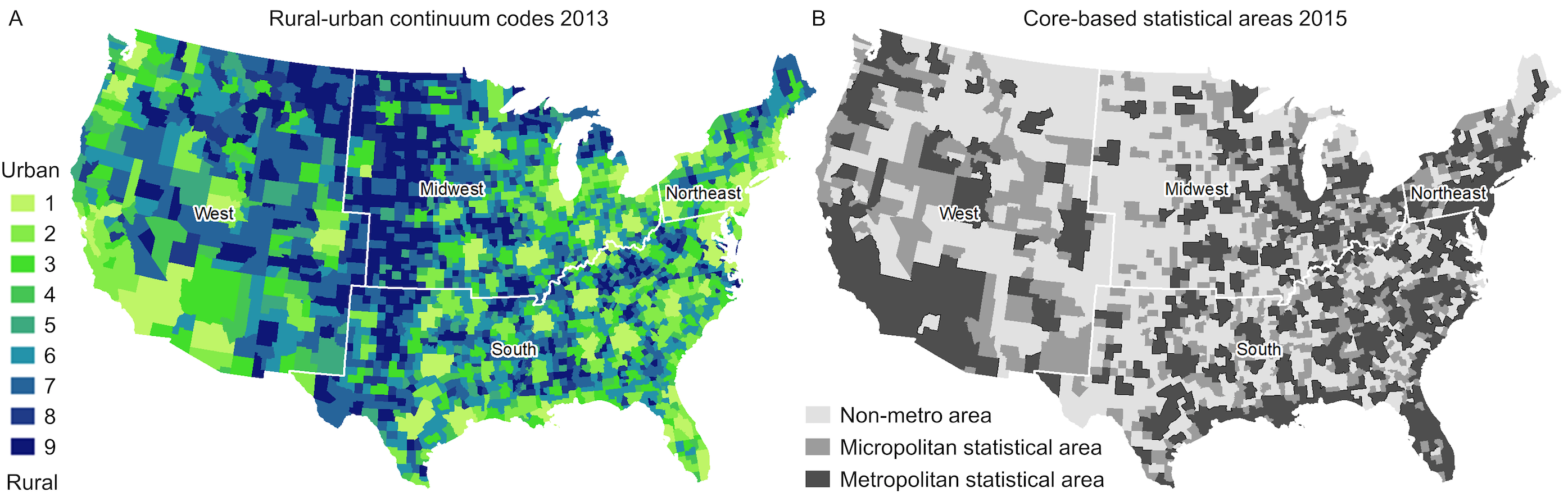}
    \caption{Rural-urban continuum and types of CBSAs. (a) Rural-urban continuum (b) CBSAs with dark gray corresponding to MSAs, medium gray corresponding to $\mu$SAs, and light gray corresponding to rural regions with neither type of CBSA. 
    }
    \label{fig:FigS1}
\end{figure*}

  \begin{figure*}[tbh!]
    \centering
    \includegraphics[width=\linewidth]{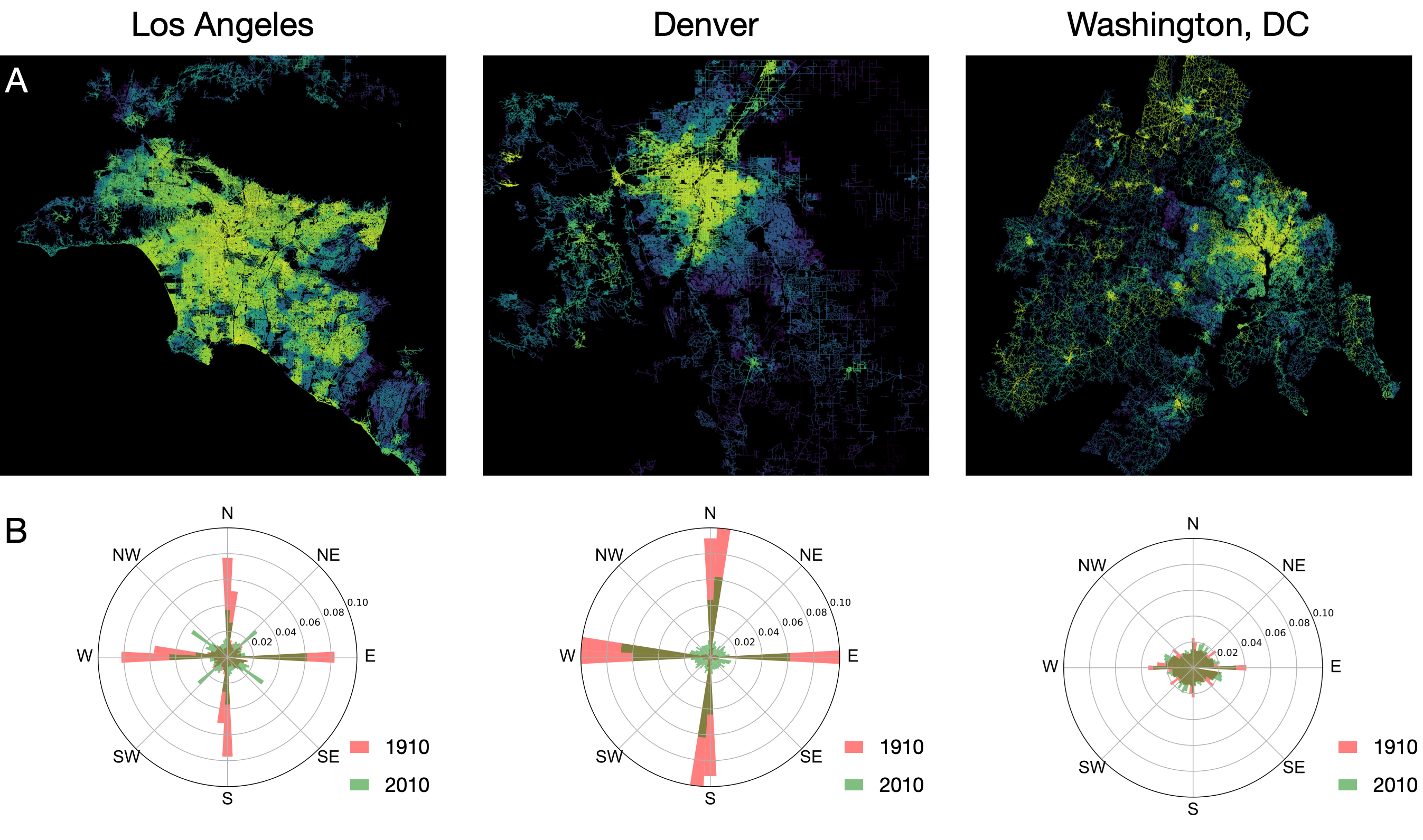}
    \caption{Spatiotemporal data. (a) Reconstructed road network over time for Los Angeles, Denver, and Washington DC metropolitan statistical areas. (b) Angle variability of roads as of 1910 and 2010
    }
    \label{fig:FigS2}
\end{figure*}

  \begin{figure*}[tbh!]
    \centering
    \includegraphics[width=0.7\linewidth]{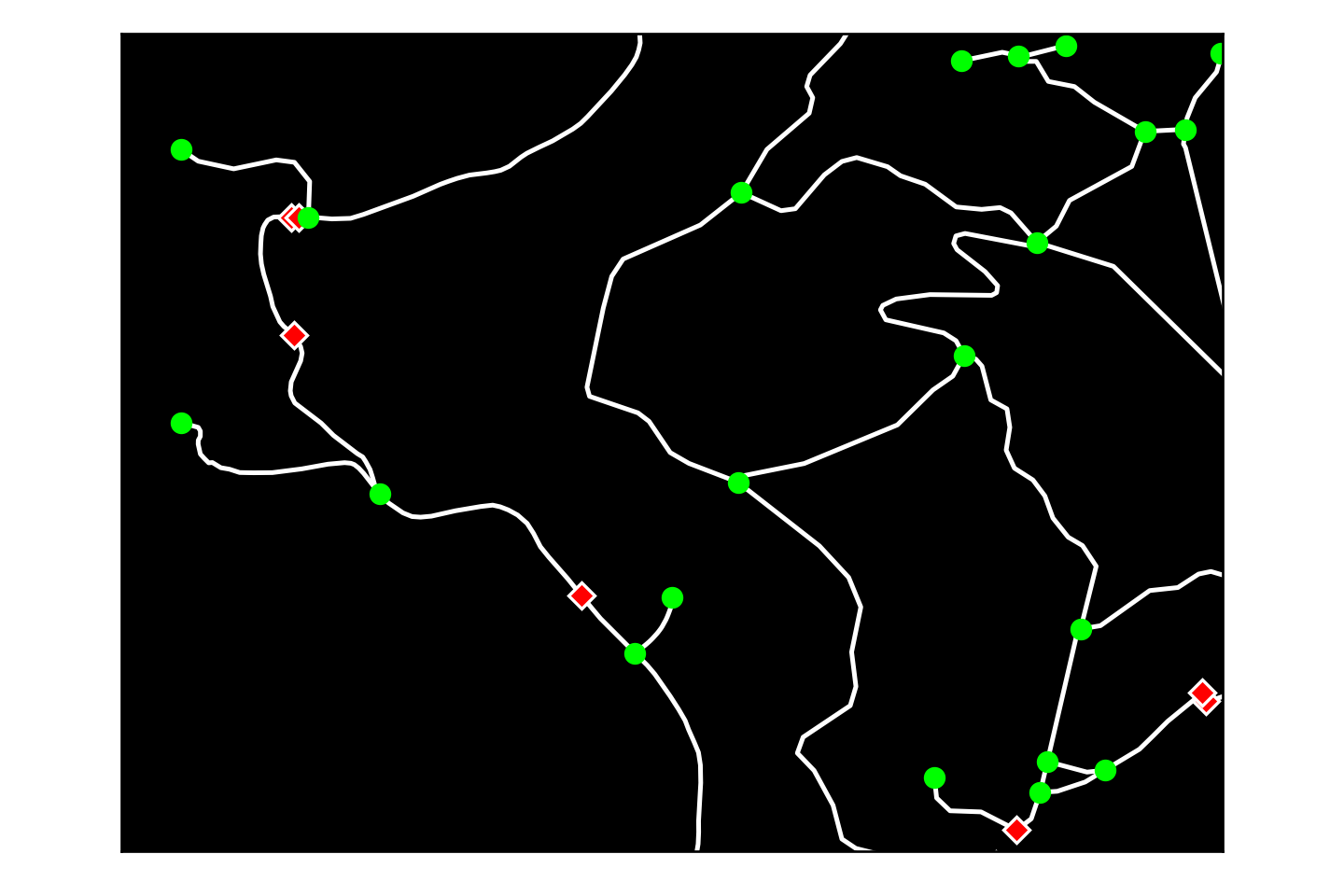}
    \caption{Degree-two nodes among roads in the Aberdeen, Washington micropolitan area as of 2015. We remove degree-two nodes from most analyses. This is because degree-two nodes are not part of an intersection. For example, near Aberdeen, Washington, we see numerous examples of degree-two nodes that form part of a continuous road (diamonds). In contrast, degree one, three, and four nodes (circles) represent valid intersections. While we use these degree-two nodes to calculate orientation entropy and or local griddedness (as their sharp angle matters in these cases), we remove such nodes from calculation of other statistics. 
    }
    \label{fig:FigS3}
\end{figure*}

  \begin{figure*}[tbh!]
    \centering
    \includegraphics[width=\linewidth]{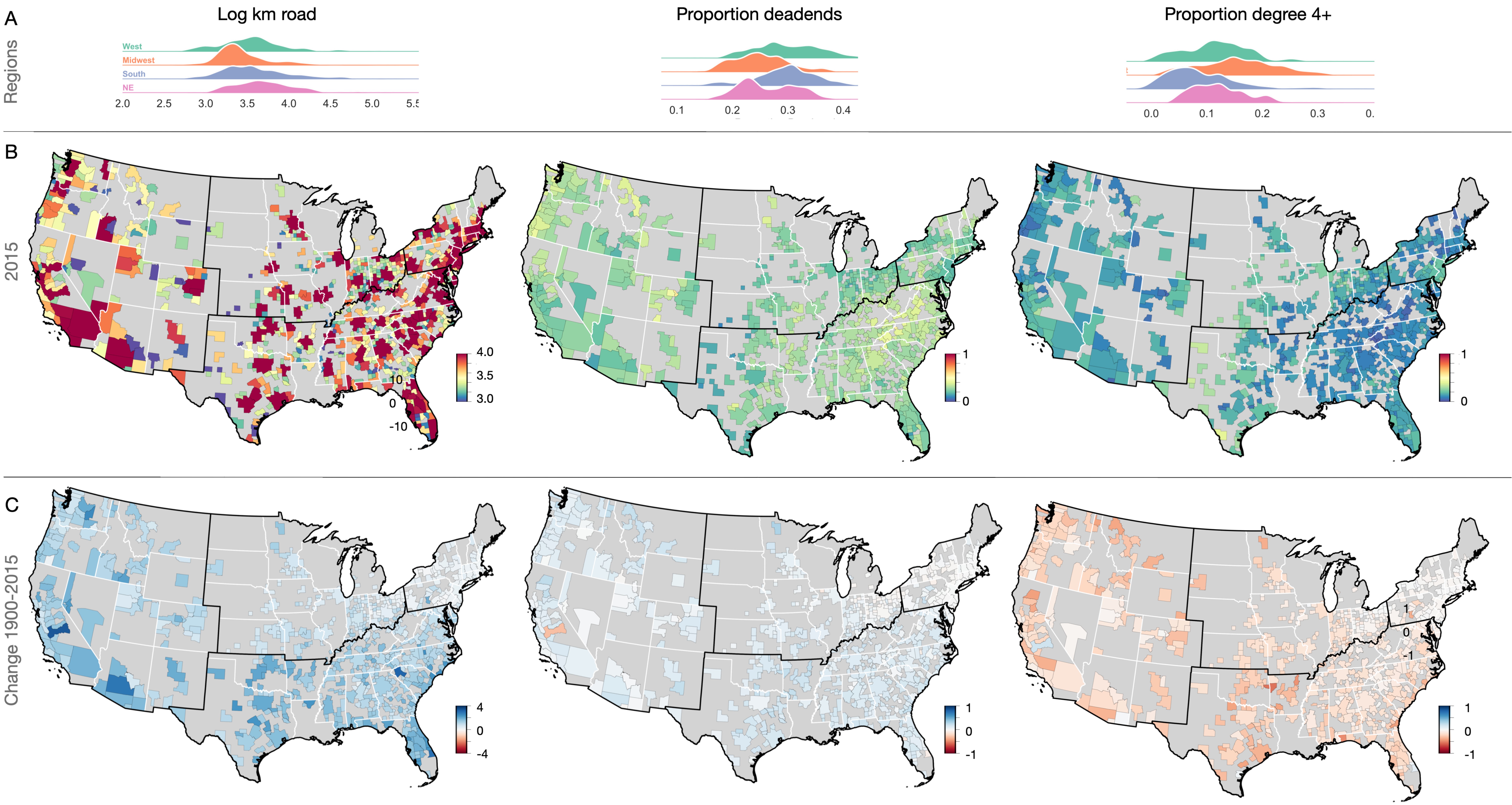}
    \caption{Other spatial statistics. We show km road, proportion of deadends, and proportion of intersections with four or more roads (a) split by regions, (b) as of 2015 for each metropolitan area, and (c) we see the change in statistics since 1900. Compare to main text Fig. 6. Most roads are in the major cities, but fewer roads have been constructed in the Northeast since 1900. The proportion of deadends is lower, and the proportion of roads with degree four or more is higher, in the Midwest. These results are consistent with the Midwest having more gridiron-like networks.
    }
    \label{fig:FigS4}
\end{figure*}

  \begin{figure*}[tbh!]
    \centering
    \includegraphics[width=\linewidth]{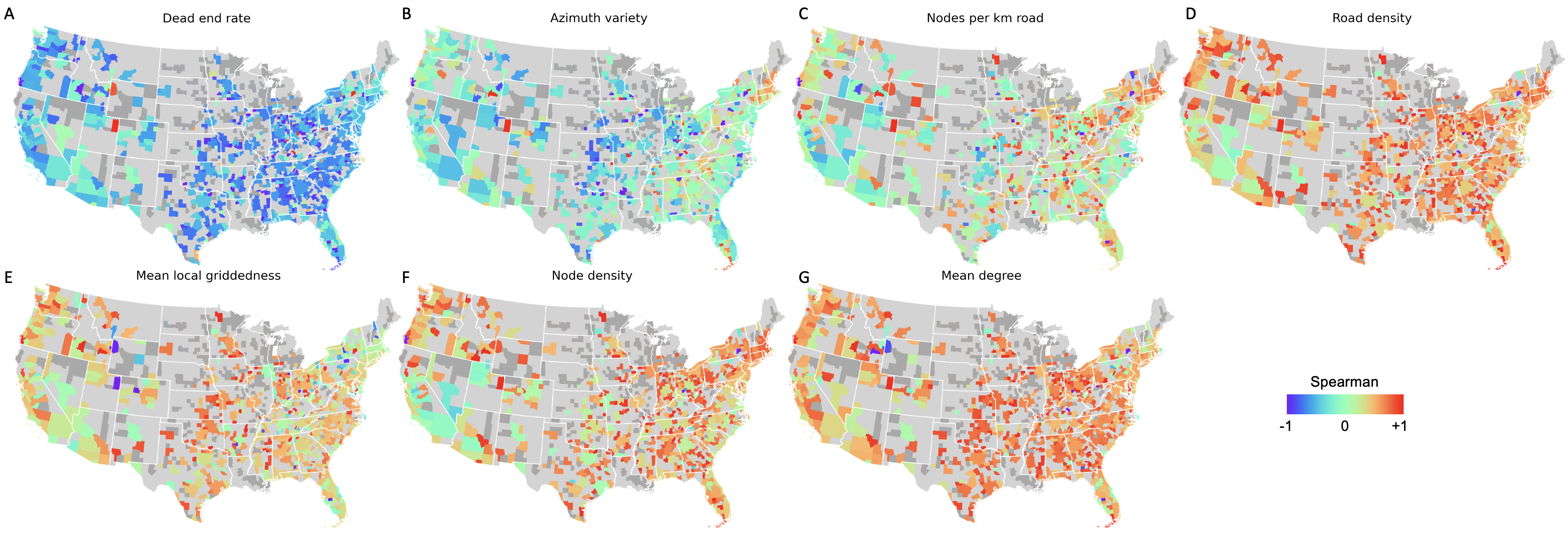}
    \caption{Spearman rank correlations between network statistics and time. Data are for dead end rate, azimuth variety, nodes per kilometer of road, road density [km$^{-1}$], mean local griddedness, node density [km$^{-2}$] and mean degree. Network statistics are calculated for roads built in a given year and are not cumulative. Light gray regions are not part of CBSAs while dark gray regions are CBSAs without enough geographic coverage (values $\le$ 40\%) or temporal completeness (values $\le 60\%$).
    }
    \label{fig:FigS5}
\end{figure*}

  \begin{figure*}[tbh!]
    \centering
    \includegraphics[width=0.9\linewidth]{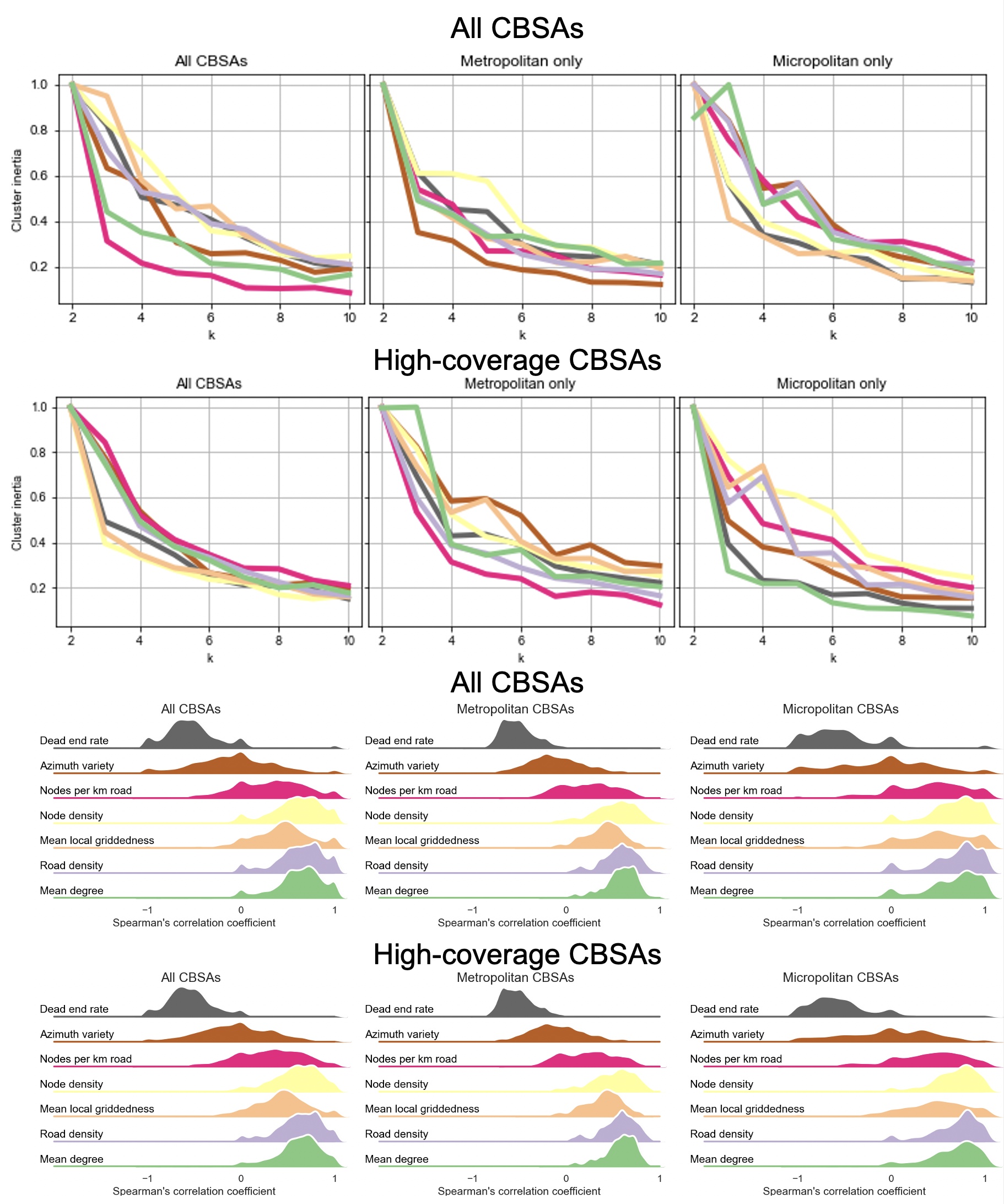}
    \caption{Clustering all and high-coverage CBSAs. Cluster inertia for (top panel) for all CBSAs, MSAs, and µSAs and (middle panel) those with $> 40\%$ geographic and $> 60\%$ temporal completeness. Each line corresponds to different statistics: dead end rate, azimuth variety, nodes per km road, node density, local griddedness, road density, and mean degree. Color legend is in the bottom two rows. Bottom two rows show ridge plots of the distribution of Spearman rank correlations between each statistic and estimated year a road was constructed for each CBSA. All findings are broadly consistent across MSAs and $\mu$SAs, as well as with high and low-coverage data. For example, we see an elbow in the inertia plot when the number of clusters, k, is 3–4. We show a detailed sensitivity analysis of these elbow plots as a function of temporal completeness and geographic coverage for azimuth variety and local griddedness in Supplementary Fig.~\ref{fig:FigS10}. In addition, we notice the dead-end rate tends to decrease in time, azimuth variety has a wide variance in its trend over time, and all other statistics tend to increase in time.
    }
    \label{fig:FigS6}
\end{figure*}

  \begin{figure*}[tbh!]
    \centering
    \includegraphics[width=\linewidth]{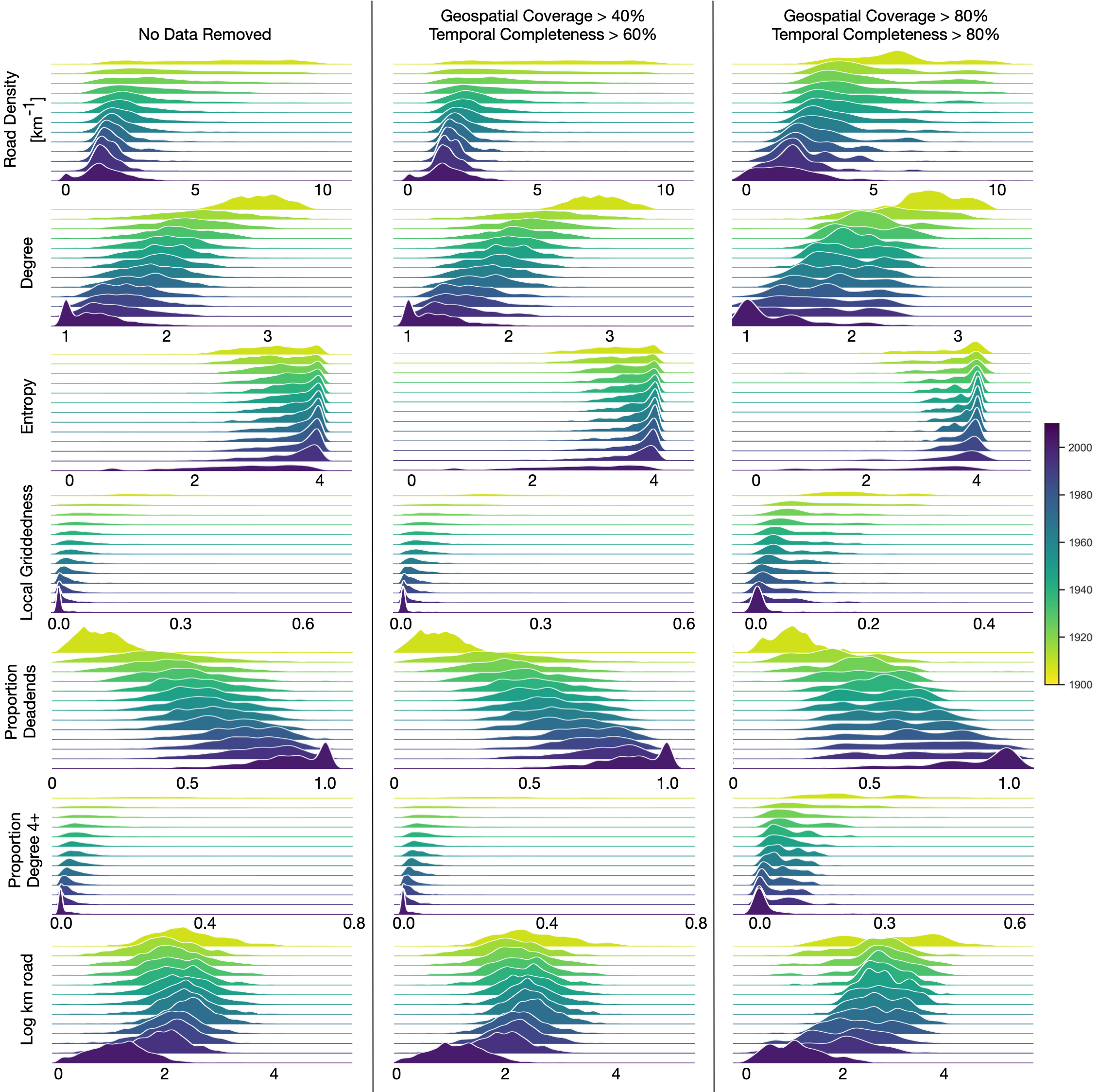}
    \caption{Ridge plots of road network statistics over time. (Left column) complete data, (middle column) main text data with geographic coverage $> 40\%$ and temporal completeness $> 60\%$, and (right column) high quality data with geographic and temporal completeness $> 80\%$. Varying coverage does not significantly affect our findings. We explore, for example, temporal trends in cities when we remove no data up to a stringent case where we only keep data with $> 80\%$ temporal completeness and geographic coverage, and we find results are qualitatively, and even quantitatively similar. 
    }
    \label{fig:FigS7}
\end{figure*}

  \begin{figure*}[tbh!]
    \centering
    \includegraphics[width=\linewidth]{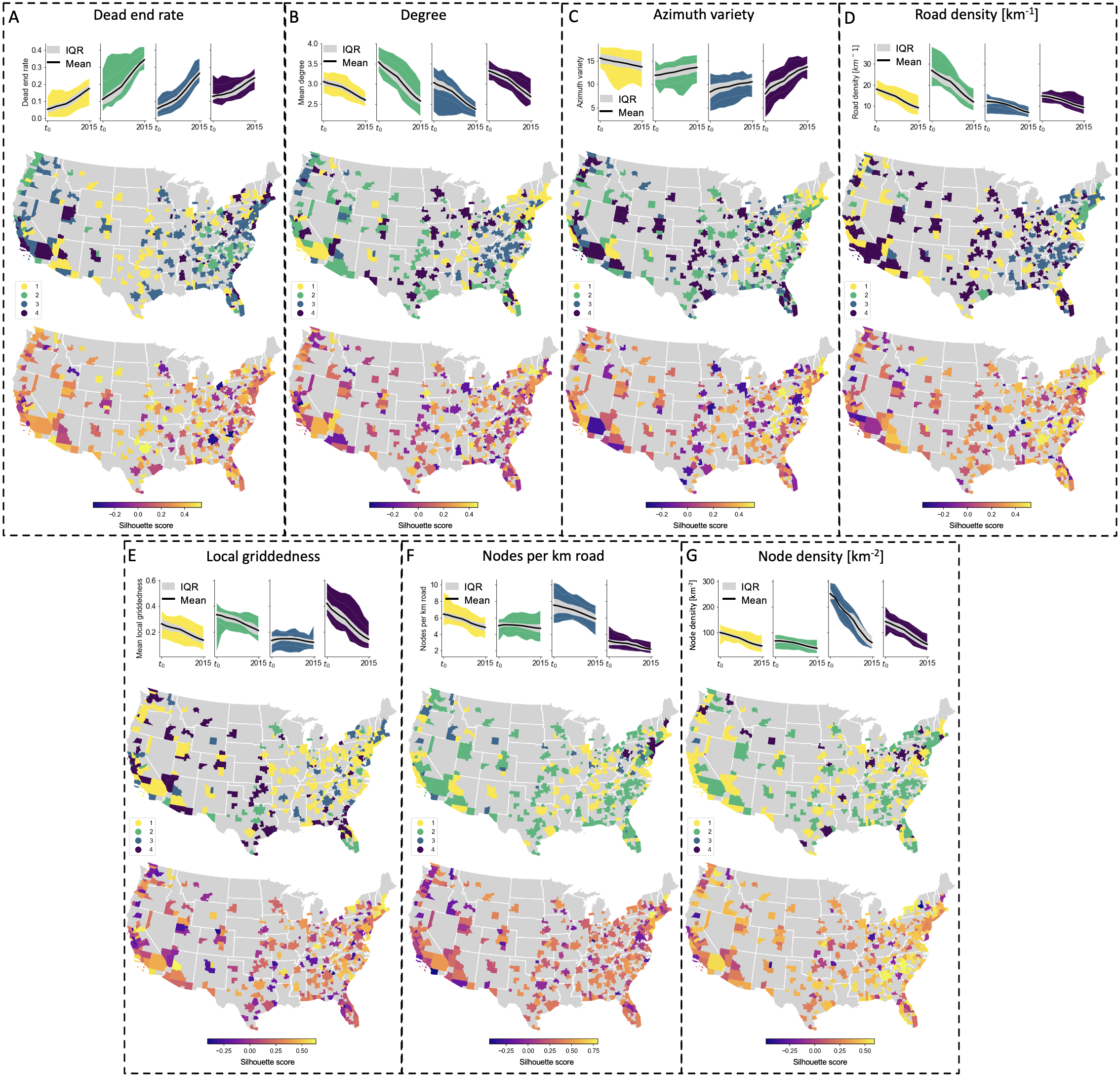}
    \caption{Clustering for seven road network statistics. (a) Dead end rate, (b) degree, (c) azimuth variety, (d) road density, (e) local griddedness, (f) nodes per km road, and (g) node density. We see significant differences in trends between clusters (top rows within subfigures), and these clusters are associated with particular regions (middle rows). Silhouette scores, which denote the degree to which cities are representative of each cluster’s mean, are shown in the bottom row.
    }
    \label{fig:FigS8}
\end{figure*}

  \begin{figure*}[tbh!]
    \centering
    \includegraphics[width=\linewidth]{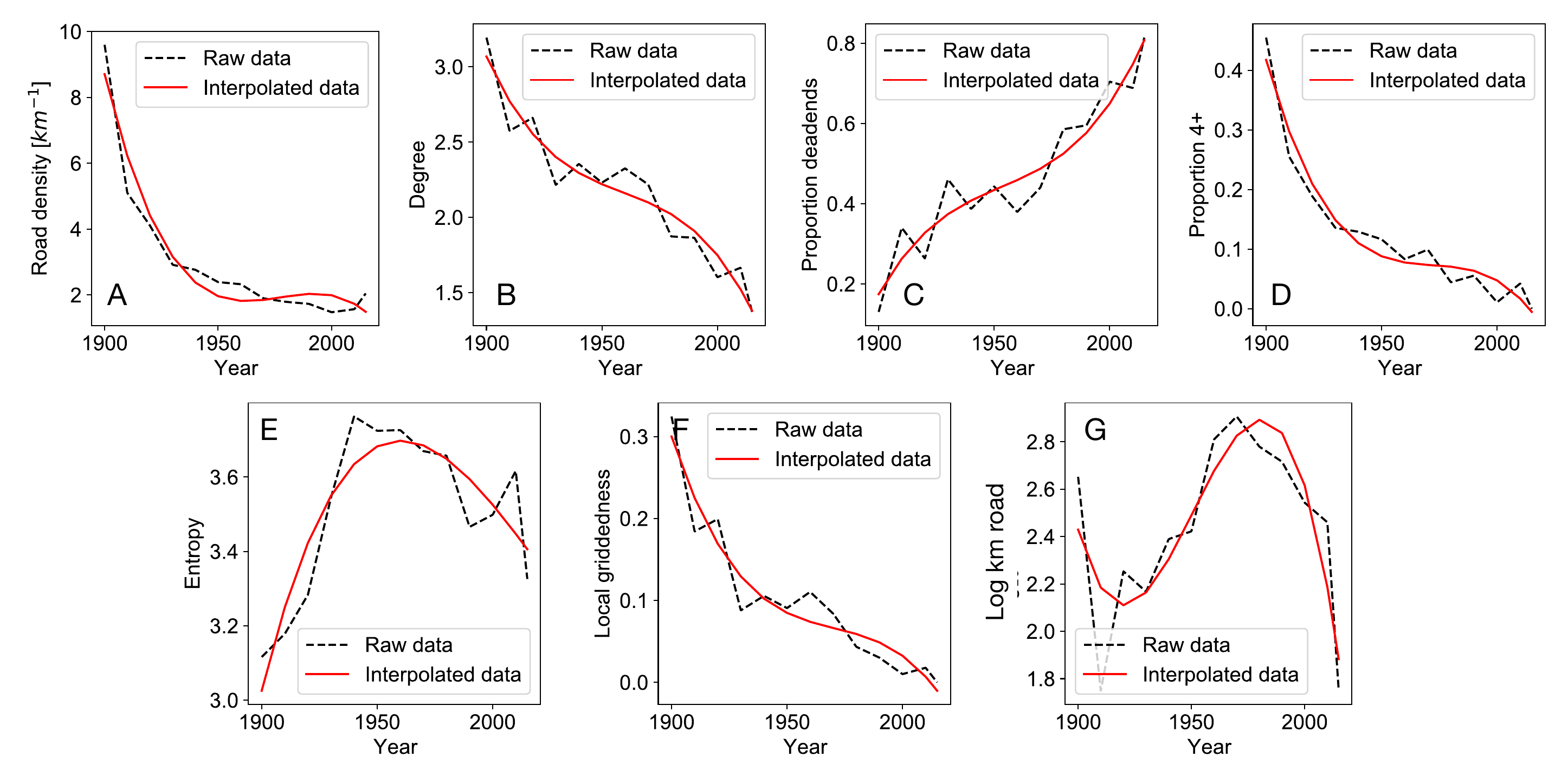}
    \caption{Interpolated data, and raw uninterpolated data, for Los Angeles MSA. We use interpolation to smooth UMAP (\cite{McInnes2018}) time series shown in main text Fig. 8. All data was centered to zero with unit variance prior to embedding. In contrast to similar visualization techniques (\cite{Uhl2021}), we smooth statistics over time before embedding to reduce jittery trends, while UMAP is used to preserve the global structure, unlike t-SNE (\cite{McInnes2018}), used previously (\cite{Uhl2021}). Interpolated data shows close agreement with the original statistics but are less noisy.
    }
    \label{fig:FigS9}
\end{figure*}

  \begin{figure*}[tbh!]
    \centering
    \includegraphics[width=0.9\linewidth]{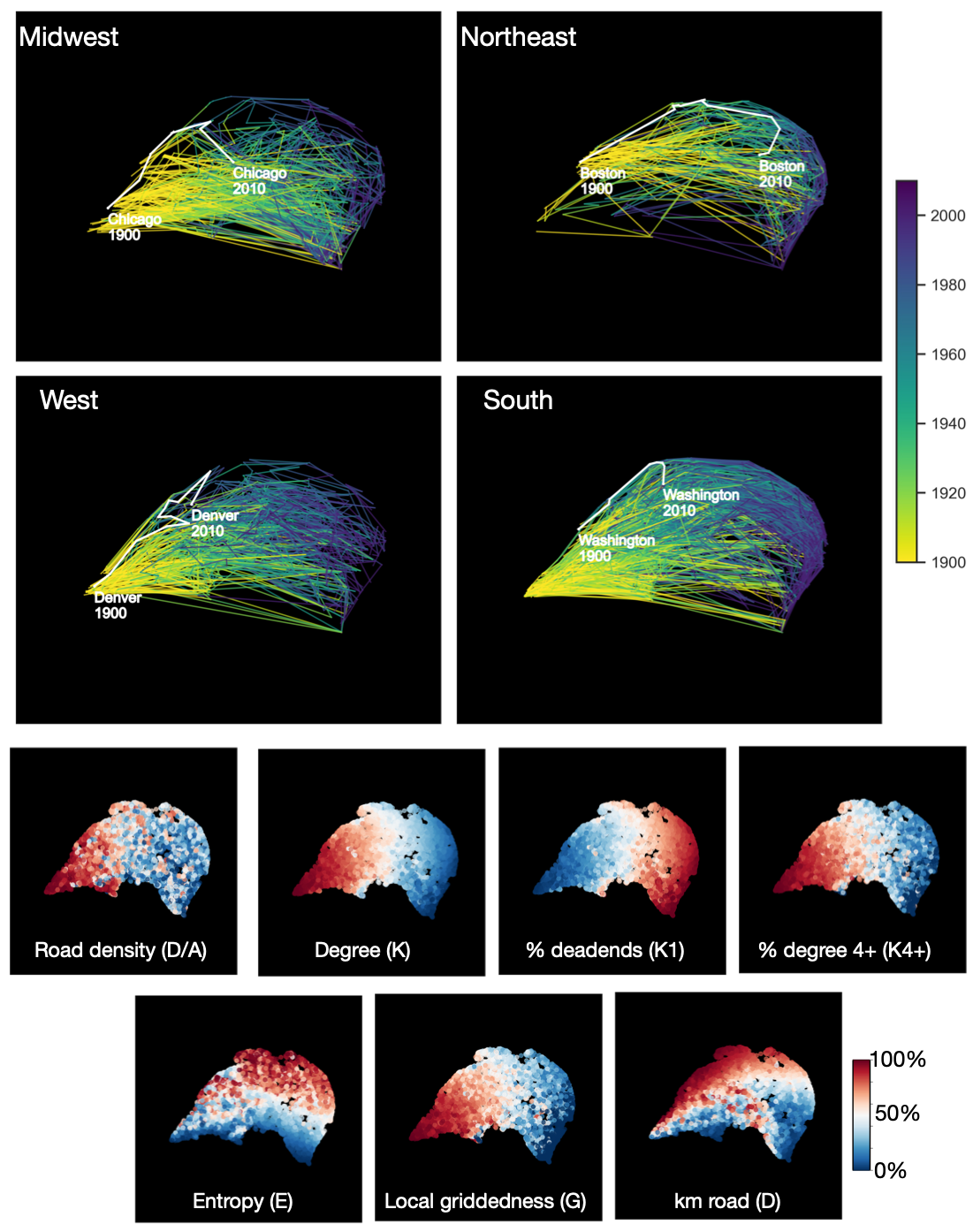}
    \caption{ UMAP embeddings of uninterpolated data. We contrast this figure (without interpolation) with main text Fig. 8 (with interpolation). If we do not interpolate between statistics, shown in Supplementary Fig.~\ref{fig:FigS12}, the UMAP plots show more noise, thus making conclusions we wish to draw more difficult to infer. 
    }
    \label{fig:FigS10}
\end{figure*}

  \begin{figure*}[tbh!]
    \centering
    \includegraphics[width=\linewidth]{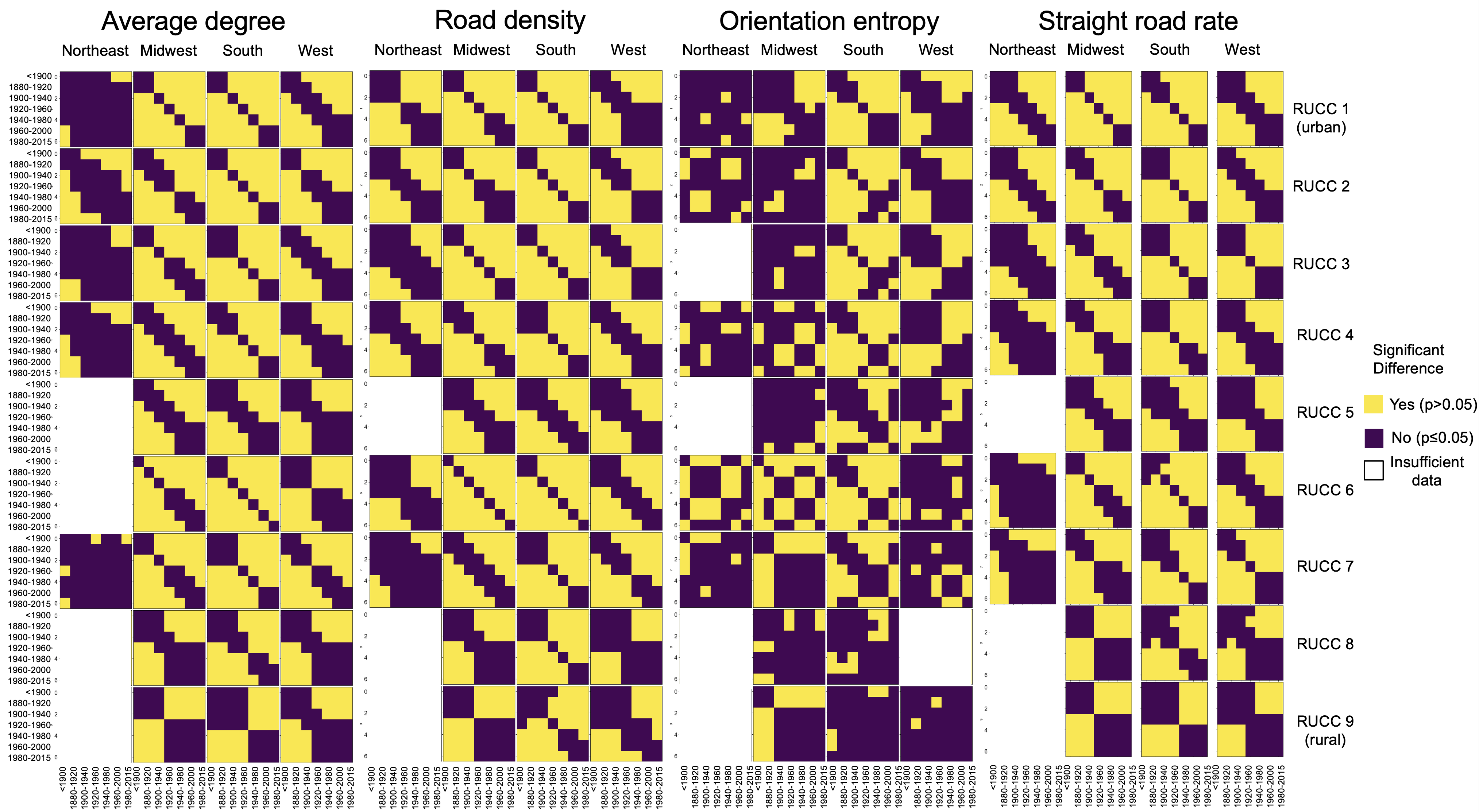}
    \caption{Statistical significance of network statistics over time. Each column are statistics constructed from road networks: average degree, road density [km$^{-1}$], azimuth (orientation) entropy, and straight road rate. X and Y-axes are these road statistics within a given time period (e.g., 1880-1920), while colors correspond to insignificant differences (dark) and statistically significant differences (light) in these statistics, based on Dunn’s test (\cite{Dunn1961}) after rejection by the Kruskal-Wallis test (p-value $< 0.05$) (\cite{Kruskal1952}). White cells contain insufficient data (two or fewer groups in a stratum.). 
    }
    \label{fig:FigS11}
\end{figure*}

  \begin{figure*}[tbh!]
    \centering
    \includegraphics[width=\linewidth]{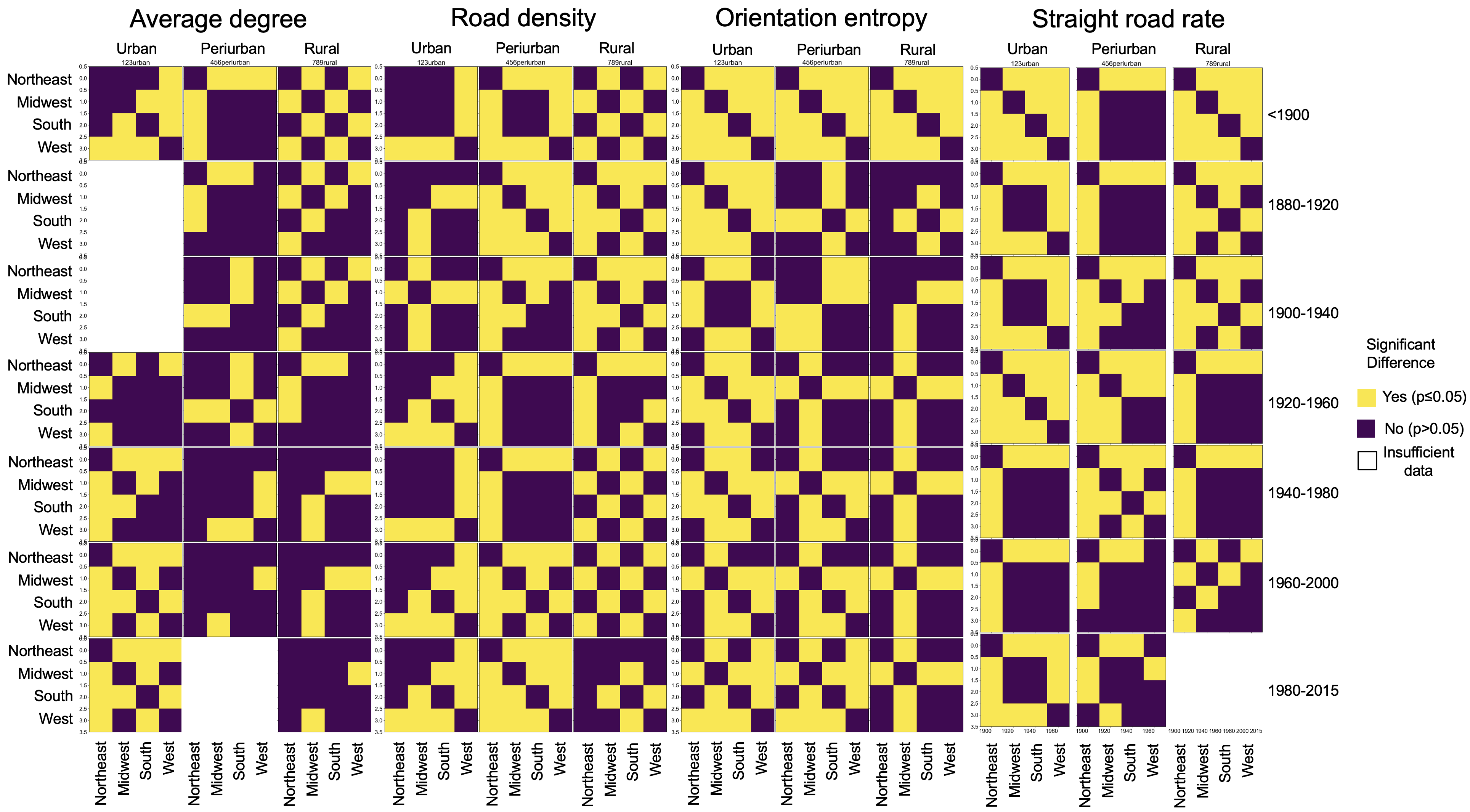}
    \caption{Statistical significance of network statistics across regions. Each column are statistics constructed from road networks: average degree, road density [km$^{-1}$], azimuth (orientation) entropy, and straight road rate. X and Y-axes are these road statistics within a given region as of 2015 (e.g., Midwest), while colors correspond to insignificant differences (dark) and statistically significant differences (light) in these statistics, based on Dunn’s test (\cite{Dunn1961}) after rejection by the Kruskal Wallis test (p-value $< 0.05$) (\cite{Kruskal1952}). White cells contain insufficient data (two or fewer groups in a stratum.).
    }
    \label{fig:FigS12}
\end{figure*}

  \begin{figure*}[tbh!]
    \centering
    \includegraphics[width=\linewidth]{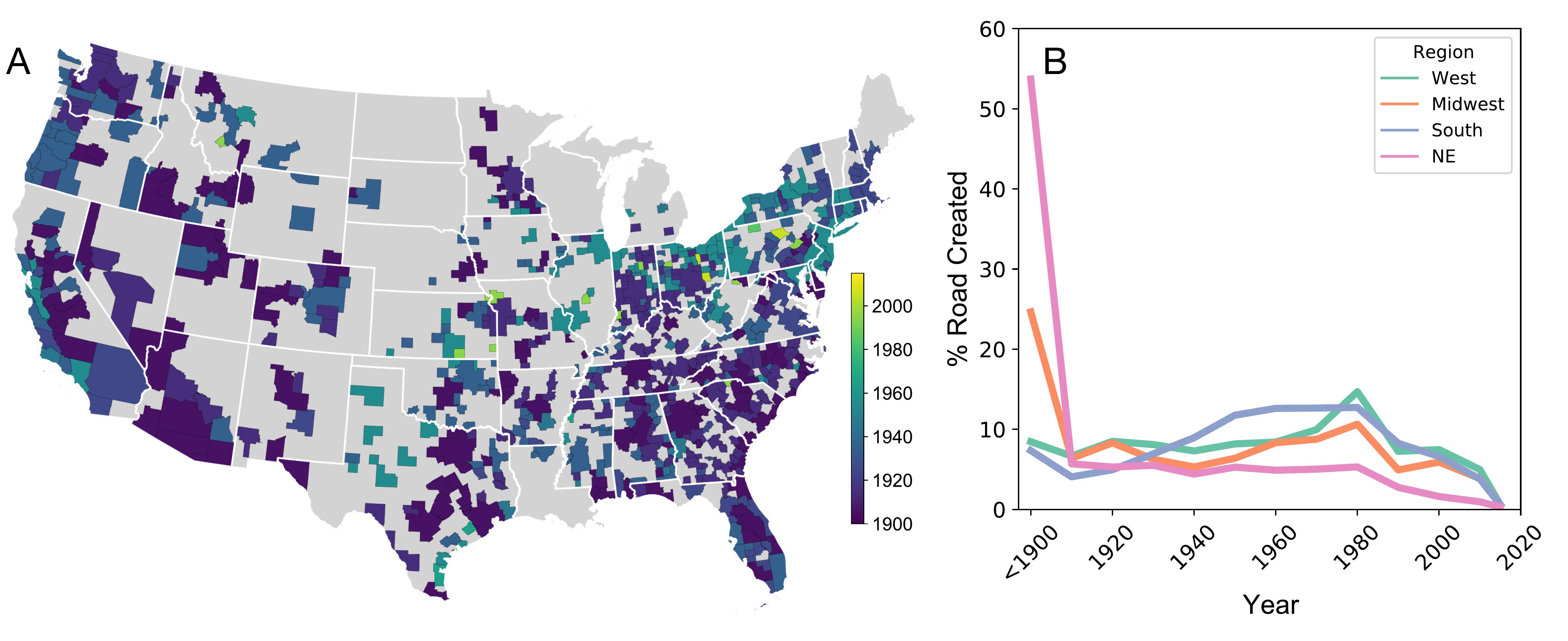}
    \caption{City growth. (a) Year of maximum development for each CBSA region, and (b) the proportion of roads constructed for each decade, split by region. We find that the year of maximum development varies across the US, with some recent growth in the Northeast and Midwest, and early growth in the South. This, however, belies that some MSAs created most of their road network before 1900. In (b), we discover that most roads were created in the Northeast by 1900. In contrast, most other regions crossed this threshold by the 1940s-1960s, in the middle of the Baby Boom. Road networks in the Northeast are therefore significantly different from those in the rest of the country, as seen in Supplementary Figs.~\ref{fig:FigS7},~\ref{fig:FigS17} \&~\ref{fig:FigS21}, and main text Fig. 6.
    }
    \label{fig:FigS13}
\end{figure*}
 
   \begin{figure*}[tbh!]
    \centering
    \includegraphics[width=0.8\linewidth]{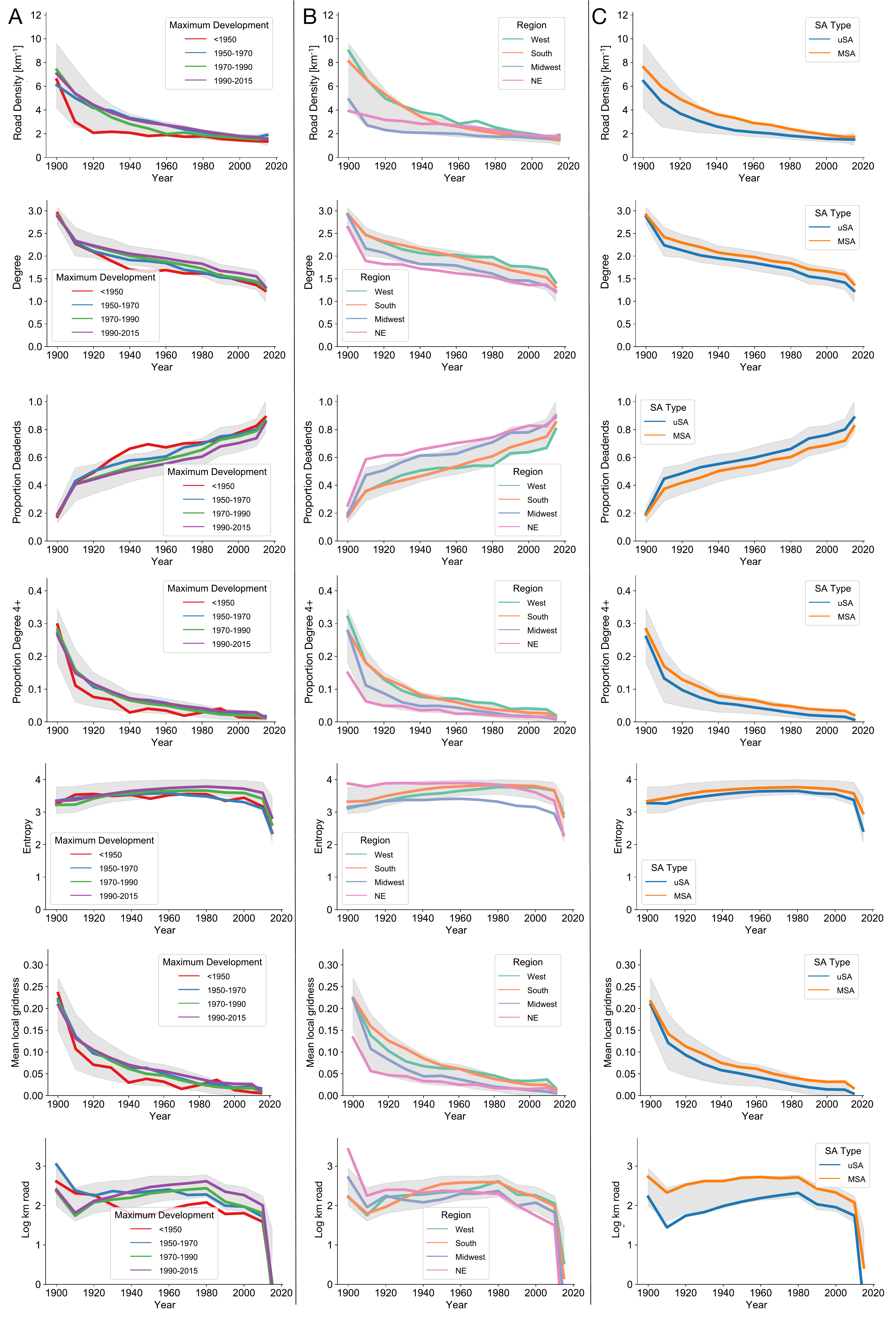}
    \caption{City statistics over time. Statistics are split by (a) year of maximum development, (b) region, and (c) whether CBSA is a metropolitan statistical area (MSA) or a smaller micropolitan area ($\mu$SA). Gray areas are interquartile ranges. The Northeast has some of the highest degree, and highest density roads, while at each decade since 1900, it trails behind other regions in the grid-like structure of new roads. These findings do not contradict with main text Figs. 6 because the Northeast made much fewer roads since 1900 compared to other regions (see Supplementary Fig.~\ref{fig:FigS20}). We also notice that the roads were most grid-like in the early twentieth century, therefore the Northeast’s road network is a historical artefact of its early construction.
    }
    \label{fig:FigS14}
\end{figure*}

 \begin{figure*}[tbh!]
    \centering
    \includegraphics[width=\linewidth]{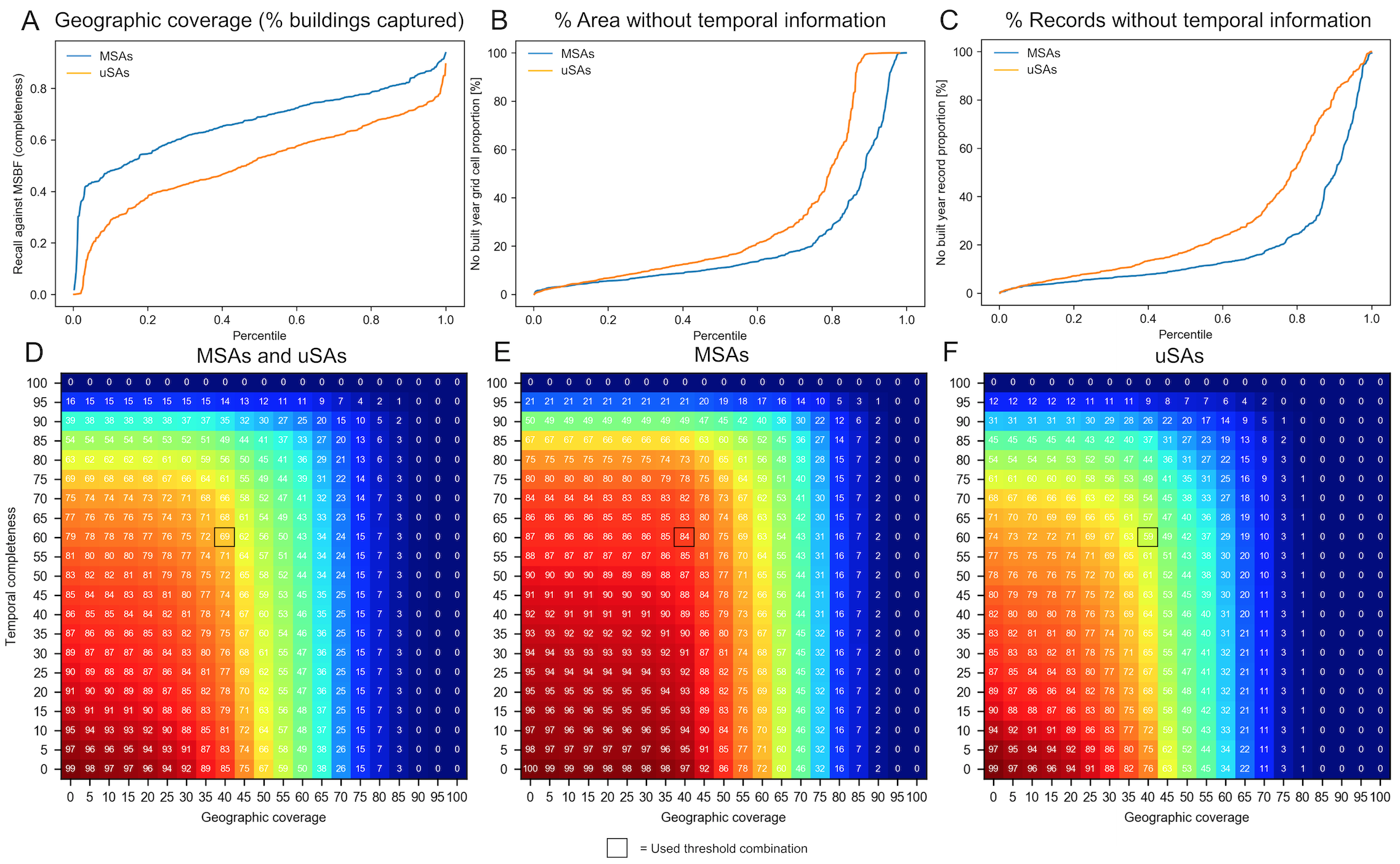}
    \caption{
    Temporal completeness and geographic coverage. (a) Percentage of buildings captured for each CBSA based on recall against Microsoft building footprints~(\cite{MBF2020}). (b) Percentage of area within each CBSA without temporal information. (c) Percentage of records without temporal information. Heatmaps of (d) MSAs and $\mu$SAs, (e) MSAs, or (f) µSAs correspond to percentage of each type of CBSA with temporal completeness (y-axis) or geographic (x-axis) coverage greater than a particular value. Black box is the threshold used in the main text. 
    }
    \label{fig:FigS15}
\end{figure*}

 \begin{figure*}[tbh!]
    \centering
    \includegraphics[width=\linewidth]{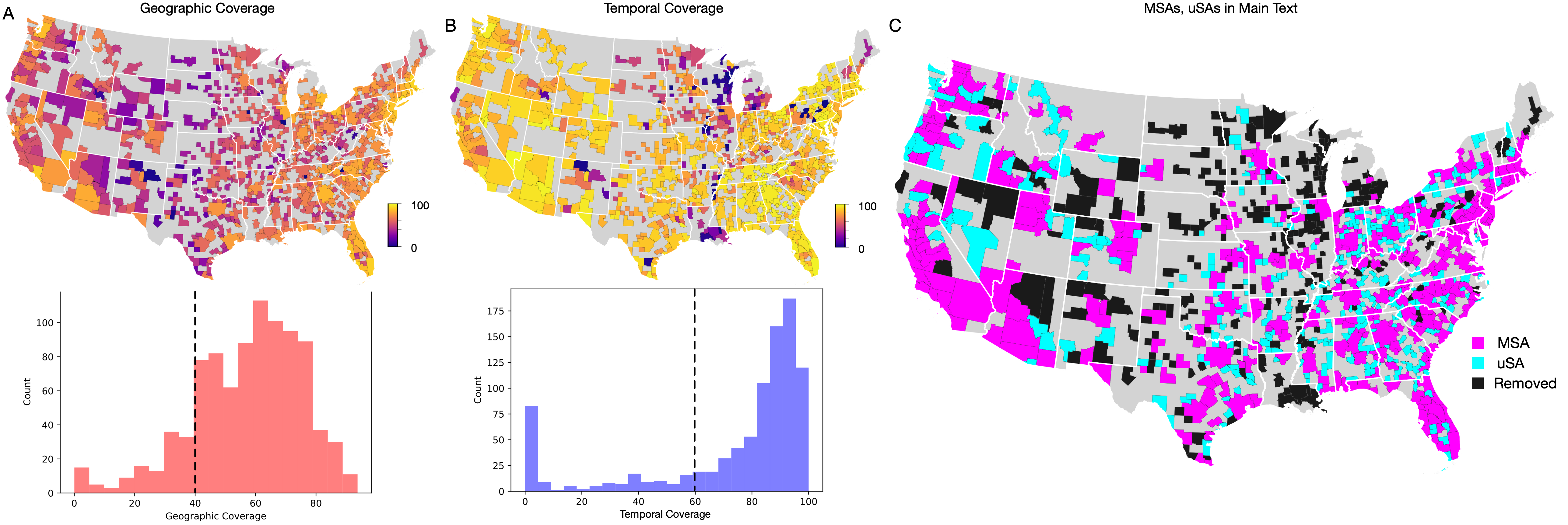}
    \caption{
    Maps and histograms of (a) geographic coverage and (b) temporal completeness across Core Based Statistical Areas (CBSAs). Dashed lines in histograms indicate minimal coverage for all data used in the main text. (c) Metro and micropolitan statistical areas (MSAs and $\mu$SAs) and data removed in main text. We show the completeness of CBSAs across the US and show a histogram of their geographic or temporal completeness. While geographic completeness is relatively uniform at around $60\%$, temporal completeness varies from nearly $0\%$ (we do not know when buildings, and therefore roads, were constructed) to nearly $100\%$ (we know when each building was constructed). Poor temporal completeness is within a few states, such as Wisconsin or Louisiana, while poor geographic coverage is more widespread, especially within the Midwest.
    }
    \label{fig:FigS16}
\end{figure*}

  \begin{figure*}[tbh!]
    \centering
    \includegraphics[width=\linewidth]{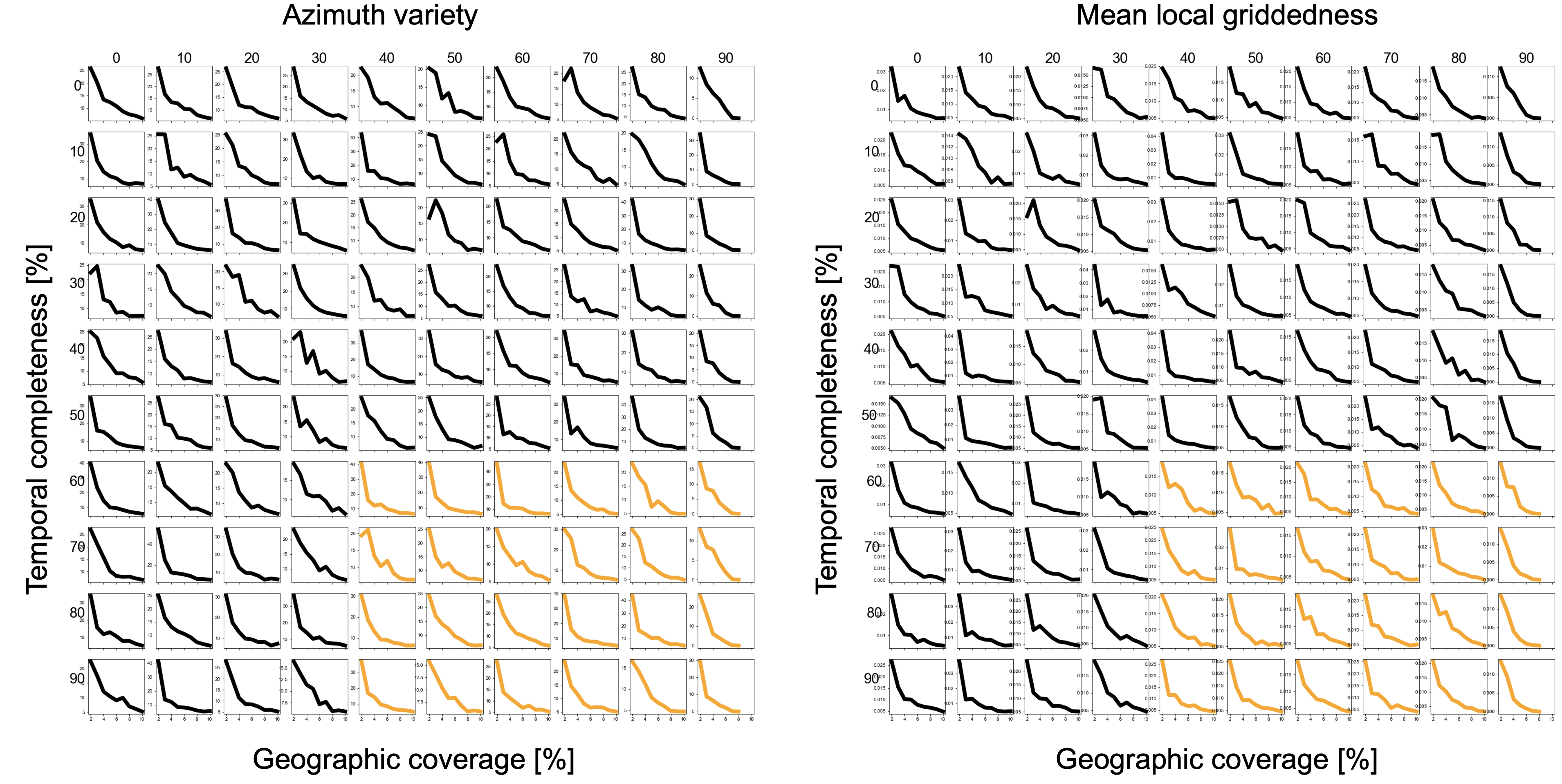}
    \caption{Inertia versus number of clusters as a function of temporal completeness and geographic coverage. Data on left is for azimuth variety, while data on the right is mean local griddedness. Comparisons of inertia plots for other statistics can be seen in Supplementary Figs.~\ref{fig:FigS9} \&~\ref{fig:FigS11}. Orange data are filtered data with temporal completeness or geographic coverage that is higher than in the main text (but is subsequently much less data).
    }
    \label{fig:FigS17}
\end{figure*}

  \begin{figure*}[tbh!]
    \centering
    \includegraphics[width=\linewidth]{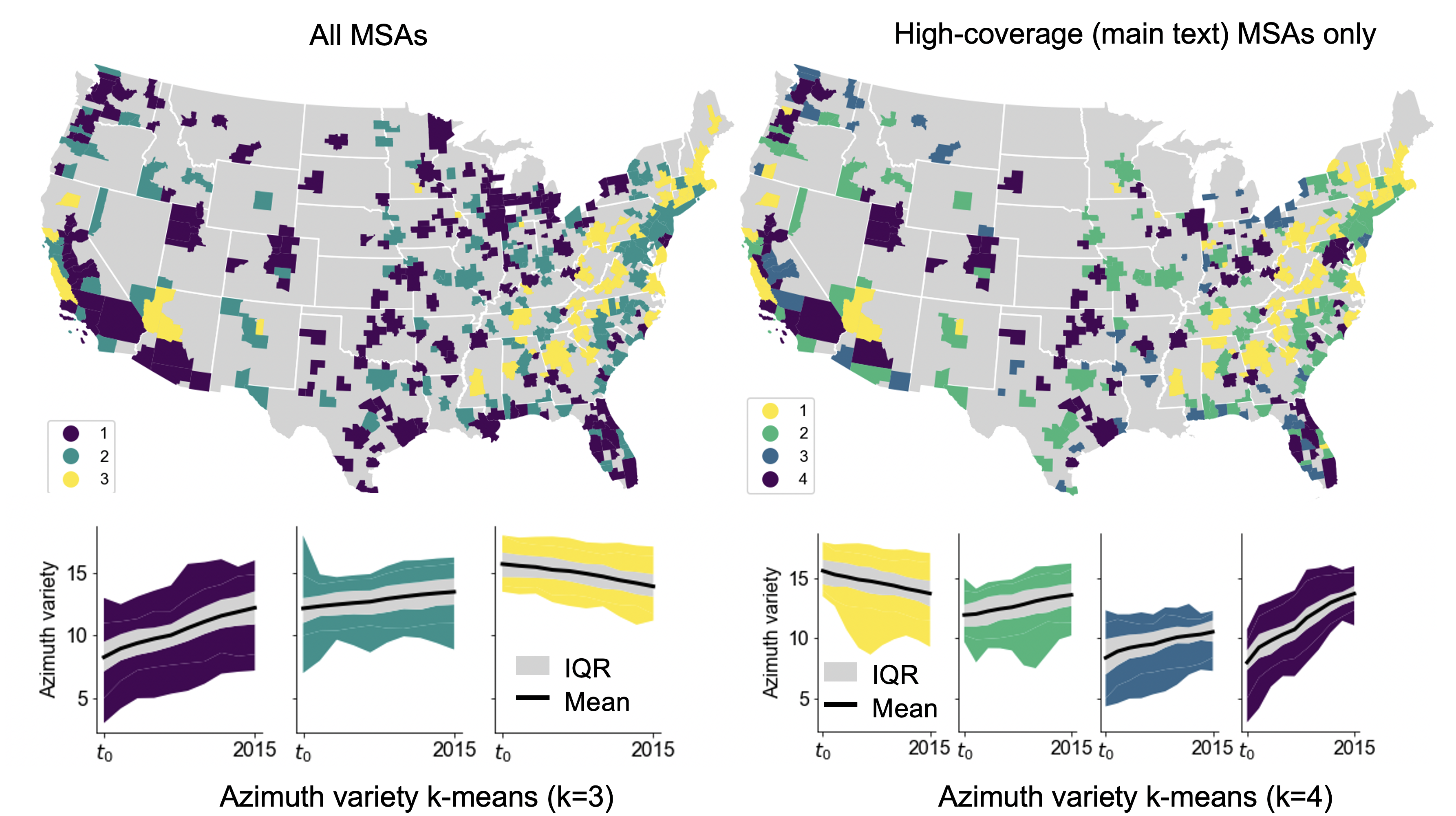}
    \caption{City evolution across the US for all MSAs and high-coverage MSAs. All MSAs (left) and high coverage MSAs (with geographic coverage $> 40\%$ and temporal completeness $> 60\%$) have similar regional clusters and each cluster evolves in similar ways. Comparisons of inertia plots for other statistics can be seen in Supplementary Fig.~\ref{fig:FigS9}. 
    }
    \label{fig:FigS18}
\end{figure*}

  \begin{figure*}[tbh!]
    \centering
    \includegraphics[width=\linewidth]{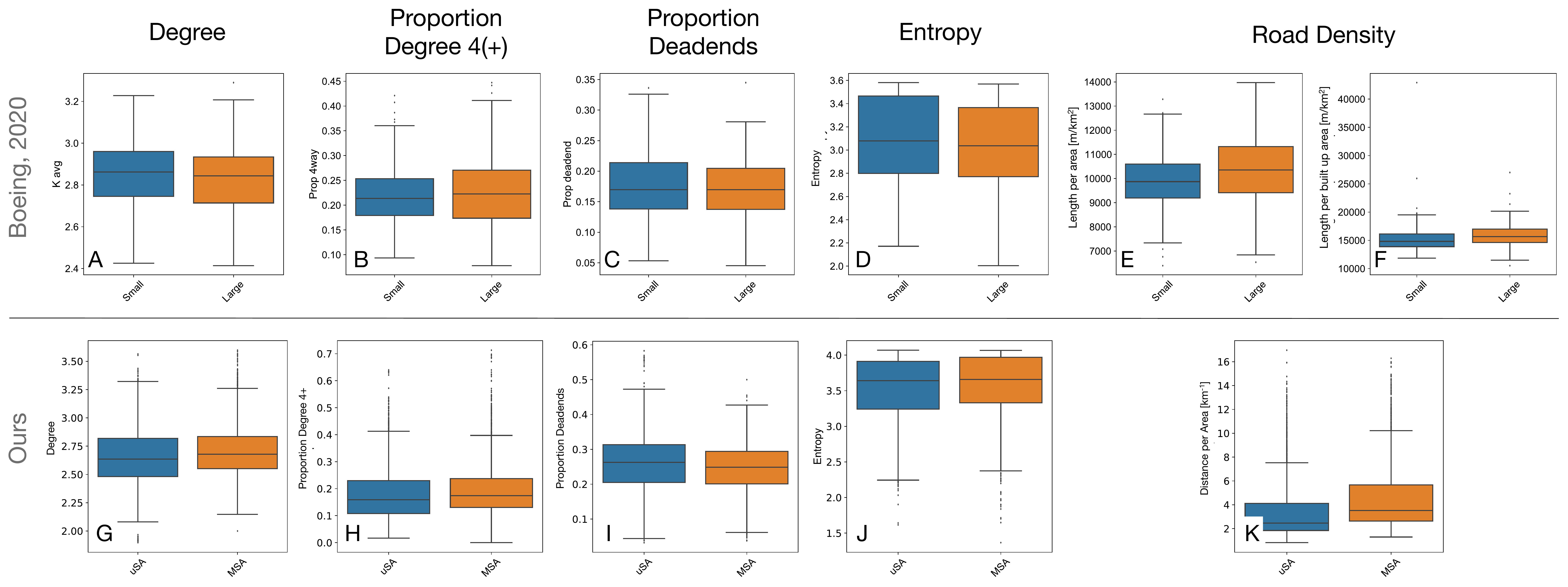}
    \caption{Data from (\cite{Boeing2020world}) and the present study split by city size. (a) Average degree, (b) proportion of four-way roads, (c) proportion of deadend roads, (d) orientation entropy, (e) road length per area, and (f) road length per built up area. Data from the present study, split by city size. (g) Average degree, (h) proportion of nodes with degree four or more, (i) proportion of deadend roads, (j) orientation entropy, (k) road length per area. Boxes within each figure correspond to interquartile ranges while whiskers are the full range barring outliers. The first row shows how larger and smaller cities, defined as those whose population is above or below 105 within the USA, have similar degree, proportion of four-way intersections and deadends, and orientation entropy. On the other hand, the median larger city has higher road densities than smaller cities. The second row shows that the present study exhibits similar qualitative findings. The smallest MSA population is 54,000 in 2015 (Carson City, NV metro area), and the largest $\mu$SA population is 220,000 (Lebanon, NH metro area) thus motivating our reasonable cut-off population of 105 when comparing large and small cities to Boeing’s study.  
    }
    \label{fig:FigS19}
\end{figure*}

  \begin{figure*}[tbh!]
    \centering
    \includegraphics[width=\linewidth]{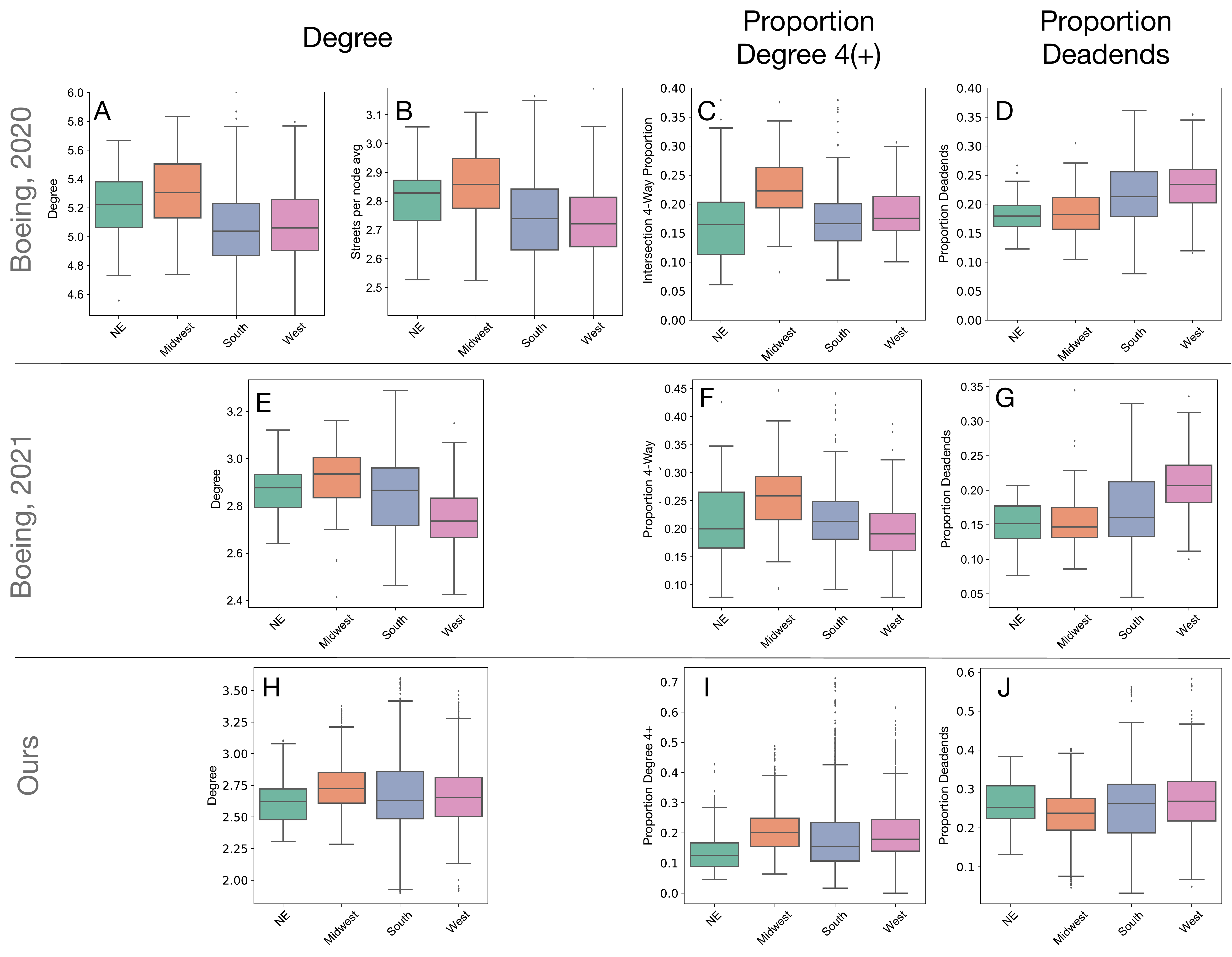}
    \caption{Data from (\cite{Boeing2020world,Boeing2020multi}), and the present study, split by region. (a) Average degree, (b) mean streets per node, (c) proportion of four-way roads, and (d) proportion of deadend roads. Data from (\cite{Boeing2020multi}) split by region: (e) average degree, (f) proportion of four-way intersections, (g) dead end proportion. Data from the present study: (h) degree, (i) proportion degree $\ge$ 4, and (j) proportion deadends. Boxes within each figure correspond to interquartile ranges while whiskers are the full range barring outliers. We find in both Boeing studies that the median Midwest city has the highest degree, 4-way intersections, and one of the lowest deadend proportions, consistent with our results in the bottom row and main text Fig. 6.
    }
    \label{fig:FigS20}
\end{figure*}

  \begin{figure*}[tbh!]
    \centering
    \includegraphics[width=\linewidth]{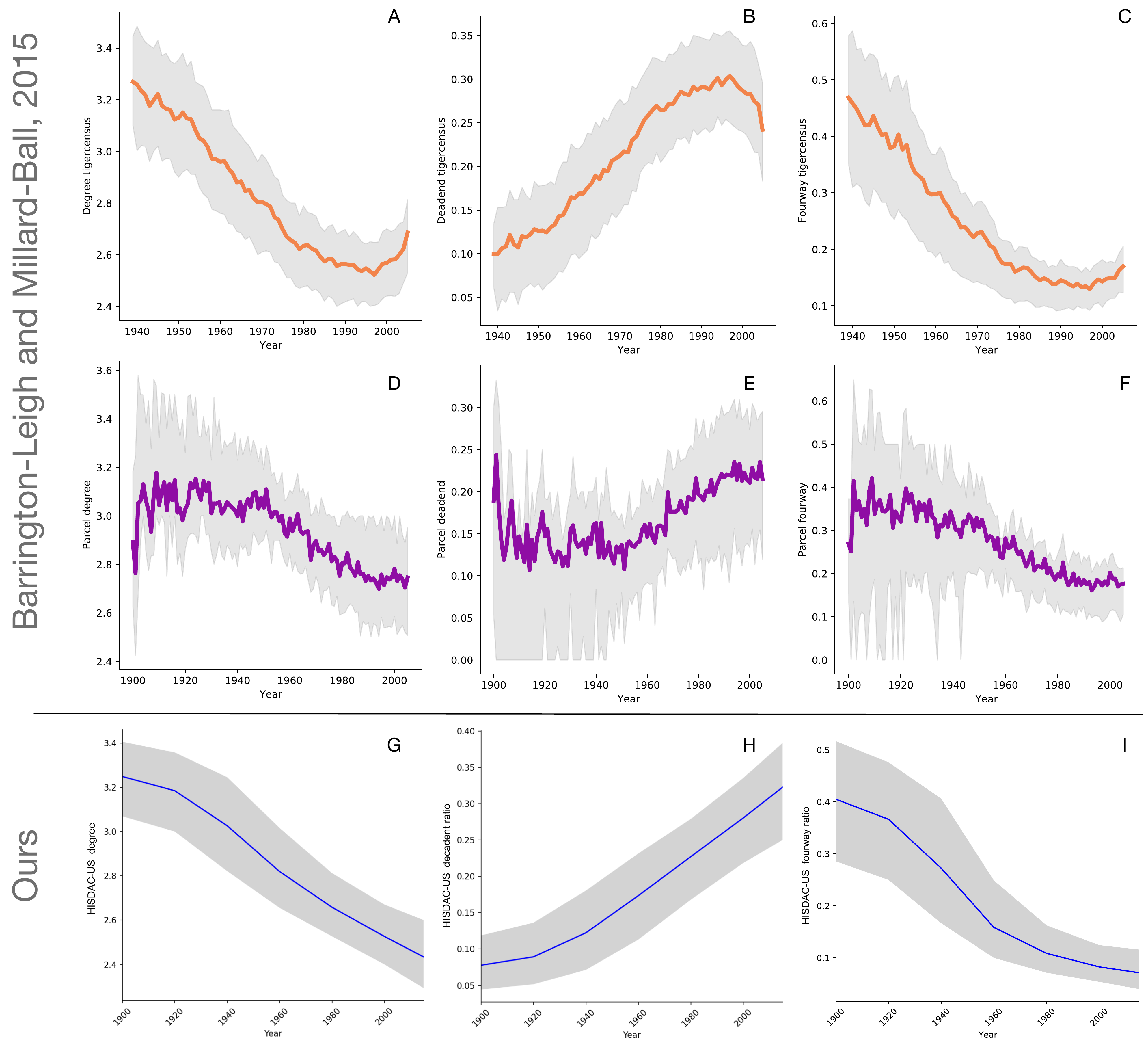}
    \caption{Comparison between present research data and data from (4). TIGER shapefiles-based (a) degree, (b) proportion deadends, and (c) proportion four-ways. We compare against parcel-based data used to reconstruct (d) degree, (e) proportion of deadends, and (f) proportion of four-ways over time. Results from the present study are shown in the bottom row for (g) degree, (h) deadend ratio, and (i) fourway ratio. Gray areas correspond to interquartile ranges. While there appears to be a change in statistic trends in (a)–(c), results in (d)–(f) show a flattening of median statistics, which is more consistent with results from the present study.
    }
    \label{fig:FigS21}
\end{figure*}

\clearpage

\end{document}